\documentclass[12pt]{spieman}  
\usepackage{amsmath,amsfonts,amssymb}
\usepackage{graphicx}
\usepackage{setspace}
\usepackage{tocloft}
\usepackage{xspace}
\usepackage{lineno}
\usepackage{color,soul}
\usepackage{xcolor}

\author[a,$^{\dagger}$]{Farzane Shirazi}
\author[b]{Ephraim Gau}
\author[b]{Md. Arman Hossen}
\author[e]{Daniel Becker}
\author[e]{Daniel Schmidt}
\author[f]{Daniel Swetz}
\author[f]{Douglas Bennett}
\author[b]{Dana Braun}
\author[g]{Fabian Kislat}	
\author[f]{Johnathon Gard}
\author[f]{John Mates}
\author[e]{Joel Weber}
\author[b]{Nicole Rodriguez Cavero}	
\author[b]{Sohee Chun}	
\author[b]{Lindsey Lisalda}
\author[b]{Andrew West}
\author[b,c]{Bhupal Dev}
\author[b,c]{Francesc Ferrer}
\author[b]{Richard Bose}
\author[e]{Joel Ullom}
\author[b,c,d,$^{\ddagger}$]{Henric Krawczynski}
\affil[a]{Clemson University, Department of Physics and Astronomy, 
Clemson University
118 Kinard Laboratory
Clemson, SC 29634}
\affil[b]{Washington University in St. Louis, Physics Department, 
1 Brookings Dr., CB 1105, St. Louis, MO 63105}
\affil[c]{McDonnell Center for the Space Sciences at Washington University in St. Louis}
\affil[d]{Center for Quantum Sensors at Washington University in St. Louis}
\affil[e]{University of Colorado,
325 Broadway St., MS 817.03,
Boulder, CO 80305-3337}
\affil[f]{National Institute of Standards and Technology (NIST), Boulder, CO 80305, USA}
\affil[g]{University of New Hampshire, Department of Physics and Astronomy and Space Science Center, 8 College Rd, Durham, NH 03824}

\cftpagenumbersoff{figure}
\cftpagenumbersoff{table} 
\begin{document} 

\newcommand{\ti}{$^{44}$Ti\xspace}
\newcommand{\casA}{Cas~A\xspace}
\newcommand{\sh}{SLEDGEHAMMER\xspace}

\input{journals.inp}
\title{The {\protect\it 511-CAM} Mission: A Pointed 511\,keV Gamma-Ray Telescope with a Focal Plane Detector Made of
Stacked Transition Edge Sensor Microcalorimeter Arrays}

\maketitle

\begin{abstract}
The 511 keV $\gamma$-ray emission from the galactic center region may fully or partially originate from
the annihilation of positrons from dark matter particles with electrons from the interstellar medium. Alternatively, the positrons could be created by astrophysical sources, involving exclusively standard model physics. 
We describe here a new concept for a 511\,keV mission called
{\it 511-CAM} (511 keV gamma-ray CAmera using Micro-calorimeters)
that combines focusing $\gamma$-ray optics with a stack of 
Transition Edge Sensor (TES) microcalorimeter arrays in the focal plane. The {\it 511-CAM} detector assembly has a projected 511 keV energy resolution of 390\,eV Full Width Half Maximum (FWHM) or better, and improves by a factor of at least 11 on the performance of state-of-the-art Ge-based Compton telescopes. 
Combining this unprecedented energy resolution with sub-arcmin angular resolutions afforded by Laue lens or channeling optics could make substantial contributions toward identifying the origin of the 511 keV emission through discovering and characterizing point sources and measuring line-of-sight velocities of the emitting plasmas. 
\end{abstract}


{\noindent \footnotesize\textbf{$^{\dagger}$}Farzane Shirazi,  \linkable{fshiraz@clemson.edu} }
{\noindent \footnotesize\textbf{$^{\ddagger}$}Henric Krawczynski,  \linkable{krawcz@wustl.edu} }


\section{Introduction}
\label{sect:intro}  
The center of our Milky Way galaxy exhibits a strong 
511 keV $\gamma$-ray signal from a yet-to-be-identified origin
\cite{1997ApJ...491..725P,2005A&A...441..513K,2005MNRAS.357.1377C,2008Natur.451..159W,2010ApJ...720.1772B,2011MNRAS.411.1727C,2020ApJ...895...44K,2020ApJ...897...45S}.
The emission may be produced by the annihilation of 
positrons from dark matter decays with electrons from the interstellar medium \cite{2018MNRAS.479.2229C}. 
Alternatively, the electrons and positrons might be created 
in X-ray binaries or other astrophysical sources, 
involving exclusively standard model physics \cite{2011RvMP...83.1001P}. 
The origin of the emission can be identified by mapping its spatial 
distribution, possibly discovering point sources, and by 
measuring the line-of-sight velocity dispersion of the 
electron-positron centers of mass immediately 
preceding the annihilation interactions \cite{2011RvMP...83.1001P}.
Two experimental approaches can be used for this purpose: 
Compton telescope observations with a
large field of view \cite{2020ApJ...895...44K,2020ApJ...897...45S}, 
and pointed experiment observations with a narrow field of view \cite{2005ExA....20..253V,barr,2014RScI...85h1101S,2017SPIE10566E..03B}.
While the first approach excels at mapping the 
large-scale distribution of the 511 keV emission,
the latter achieves far superior angular and energy resolutions.
The importance of understanding the 511 keV emission makes a case for implementing both approaches.

In this paper, we describe the new {\it 511-CAM} mission concept
that combines imaging or concentrating $\gamma$-ray optics with novel focal plane instrumentation. Even though pointed
$\gamma$-ray missions have been proposed in the past \cite{2005ExA....20..253V,barr,2014RScI...85h1101S,2017SPIE10566E..03B}, our concept leverages, for the first time, the energy resolutions of microcalorimeters
such as the \sh ({\bf S}pectrometer to {\bf L}everage {\bf E}xtensive {\bf D}evelopment of {\bf G}amma-ray T{\bf E}Ss for {\bf H}uge {\bf A}rrays using {\bf M}icrowave {\bf M}ultiplexed {\bf E}nabled {\bf R}eadout) array developed by the NIST quantum sensor group.
The \sh detectors achieve an unprecedented energy resolution of 
$55$ eV FWHM at 97 keV
\cite{2012RScI...83i3113B,noroozian_etal_2013,mates_etal_2017}.
{The {\it 511-CAM} based on the \sh technology is expected to 
give a 511~keV energy resolution of {$\le$\,330\,eV FWHM}, thus achieving 
a factor of $\ge$\,11 improvement in energy resolution compared to the current state-of-the-art HPGe (High Purity Ge) based Compton telescopes \cite{2020ApJ...895...44K}.} 
The dramatic improvement comes from the superior energy resolution for each 
individual energy deposit provided by the calorimetric approach, as well as 
the smaller number of interactions in high-$Z$ absorbers used.
The {\it 511-CAM} focal plane instrumentation can be combined 
with Laue lens imaging optics or channeling concentrator 
optics with focal lengths of 12\,m or more 
to focus $\gamma$-rays onto the {\it 511-CAM} detector assembly 
(see Section \ref{s:optics}). 
The {\it 511-CAM} concept would be well-suited for a (Ultra) 
Long Duration Balloon ((U)LDB) balloon flight or 
a satellite borne mission. 

The rest of the paper is organized as follows. 
Section \ref{s:science} highlights the scientific motivation for 
{\it 511-CAM}-type missions. Sections \ref{s:optics}, \ref{s:511-Cam}, and 
\ref{s:telescope}
describe the optical, focal plane, and telescope
design of a balloon-borne {\it 511-CAM} mission, respectively.
We report on simulations of the performance of the focal plane detector 
in Section \ref{s:perf}, and Section \ref{s:performance} presents the 
sensitivity achievable with the balloon borne mission.
Finally, we summarize and discuss the results in Section \ref{s:discussion}. 
\section{Scientific Motivation}
\label{s:science}

\subsection{Identifying the Origin of the 511 keV Emission from the Galactic Center}
The {\it OSSE}, {\it INTEGRAL}, and {\it COSI} (Compton Spectrometer and Imager) missions mapped 
the bright 511\,keV emission from the galactic center region  
\cite{1997ApJ...491..725P,2005A&A...441..513K,2005MNRAS.357.1377C,
2008Natur.451..159W,2010ApJ...720.1772B,2011MNRAS.411.1727C,
2020ApJ...895...44K,2020ApJ...897...45S}. 
{For reference,} the {\it COSI} mission measured the 511\,keV line with a 
4.3\,keV FWHM energy resolution.

The review article by Prantzos et al.\,summarizes a number of potential dark matter and astrophysical explanations for the emission \cite{2011RvMP...83.1001P}. 
The dark matter explanations for the emission are placed into two categories: the decay and annihilation of $\sim$MeV particles, or 
the de-excitation of dark matter in models with hidden sectors. 
Prantzos et al.\,note that different models predict distinct spatial distributions for the resultant positrons. 
Siegert et al.\,published upper limits and possible detections
on the 511\,keV fluxes from 39 dwarf satellite galaxies.
The values found are well above the expectations from
interpreting the Milky Way emission as  purely a dark 
matter signal \cite{2016A&A...595A..25S,2017JCAP...02..049L}. Moreover, the MeV-scale dark matter annihilation interpretation of the 511-keV excess was challenged by measurements of the damping tail of the cosmic microwave background~\cite{Wilkinson:2016gsy}, but refined measurements~\cite{Sabti:2019mhn}, as well as modified dark matter annihilation mechanisms~\cite{Ema:2020fit}, still allow for this possibility. 
Cai et al.\ discuss the compatibility of mixed dark matter and primordial black hole explanations of the 511\,keV emission and the {\it INTEGRAL} results, and highlight the importance of more sensitive follow-up observations \cite{2020arXiv200711804C}.  The possibility of Hawking evaporation of a population of primordial black holes concentrated in the inner galaxy has also been considered, with a testable prediction of diffuse gamma ray emission from the inner galactic halo~\cite{Keith:2021guq}.
A {\it 511-CAM}-type mission with excellent spatial and energy resolutions can
distinguish between discrete astrophysical sources and a 
potential dark matter origin of the 511 keV excess. 
In addition, the observations can provide important constraints on different dark matter models, as well as model-independent constraints on the dark matter density profile in the inner galactic region. 
The most significant contributions of such a mission would thus be to observe the brightest 511\,keV $\gamma$-ray regions to search for bright central core emission and point sources.
In the case of a positive detection, the data 
would constrain the line-of-sight 
velocity distribution of the emitting plasma. 
\subsection{Additional Scientific Goals of Pointed Gamma-Ray Observatories}
The unprecedented energy resolution of a {\it 511-CAM}-based $\gamma$-ray mission can be used
to scrutinize various source classes for electron-positron annihilation emission.
A {\it 511-CAM} would thus enable a search for 511\,keV emission from pair plasmas in X-ray binaries such as Cyg X-1 \cite{1992MNRAS.258..657C,2015MNRAS.451.4375F}. 
The analysis of the {\it NUSTAR} observations of 
accreting stellar mass black holes indicates that 
the coronal temperatures frequently run against 
the upper limit at which pair cooling becomes
significant, indicating that pair plasma may indeed be
present. A {\it 511-CAM}-type mission 
would be particularly well-suited to finding pair plasma if the plasma is sufficiently far away from the black hole, such that the gravitational and Doppler broadening of the line is smaller than the {\it 511-CAM}'s energy resolution.

Additionally, electron-positron pairs may also be abundantly present 
in the collimated plasma outflows (jets) from stellar 
mass and supermassive black holes \cite{2018A&A...620A..41V,2019ApJ...880...37P}.
The {\it INTEGRAL}-SPI instrument indeed revealed evidence for a 511 keV emission
line in an outburst of the microquasar V404 Cygni \cite{2016Natur.531..341S}.
The detection of a 511 keV emission (possibly Doppler-shifted, owing 
to the relativistic motion of the plasma) would give us a unique probe of the
makeup and flow velocity of the jets.
High–sensitivity, excellent energy resolution observations with a {\it 511-CAM} would thus be
highly desirable.

A {\it 511-CAM} could furthermore observe X-ray lines from nucleosynthesis in various
astrophysical settings. Harrison et al.\ describe prominent lines at 158, 749, and 812 keV (from the decay of $^{56}$Ni) and at 847 and 511 keV (from the decay of $^{56}$Co) in the spectra of SNe Ia. These two decays power much or all of the emission from SNe Ia \cite{harrison_etal_2011}, and a {\it NUSTAR} follow-up mission could detect several tens of
these objects per year. A deeper understanding of SNe Ia explosion mechanisms would greatly benefit many areas of study, given their use as standard candles in astronomy and cosmology. 

Finally, such an instrument could have even more relevant objects of study: Diehl et al.\ list prominent $\gamma$-ray lines at 
68\,keV and 78\,keV (from the radioactive decay of $^{44}$Ti in core-collapse supernovae), 
at 122\,keV and 158\,keV (from decays of $^{57}$Ni and $^{56}$Ni, respectively, 
in supernova nucleosynthesis), at
478\,keV (from the decay of $^7$Be in nova nucleosynthesis), 
and at 812\,keV and 847\,keV (from the decays of $^{56}$Ni and $^{56}$Co, respectively, 
in supernova nucleosynthesis) as important diagnostics for their respective source classes \cite{2017AIPC.1852d0004D}. Similarly, the remnant of a neutron star merger that occurred between a thousand 
and a million years ago would emit prominent $\gamma$-ray lines with energies of 400 keV and 700 keV (from the decays of $^{125}$Sb, $^{126}$Sb, $^{140}$La, 
$^{213}$Bi, $^{214}$Bi, and $^{246}$Am), along with other lines at lower and higher energies
\cite{korobkin_2019}. 
However, experiments of reasonable size can detect only nearby galactic remnants.
\section{Options for the {\it 511-CAM} \texorpdfstring{$\gamma$}{Gamma}-Ray Optics}
\label{s:optics}
\subsection{Laue Lenses} 
A {\it 511-CAM} detector could be used in the focal plane of a pointed imaging or non-imaging (concentrating)
$\gamma$-ray telescope. Two concepts for focusing $\gamma$-rays onto a detector have been explored in this study: Laue lens optics and channeling optics. Both methods can be used to collect photons over an area substantially larger than the detector area.

\begin{figure}
  \centering
  \includegraphics[width=.45\textwidth]{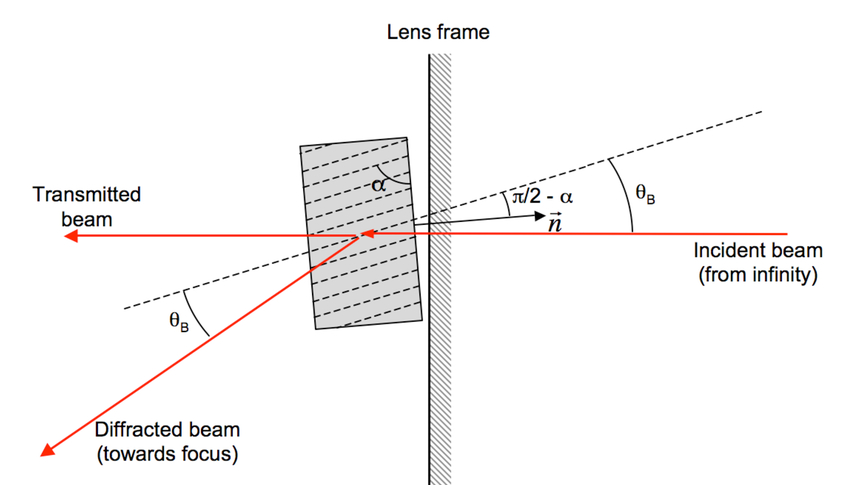}\quad%
  \includegraphics[width=.45\textwidth]{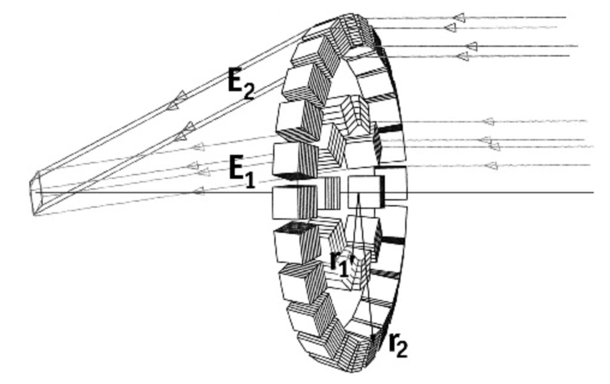}\quad%
  \caption{Working principle of the Laue lens. 
  Left: $\gamma$-rays undergo Bragg diffraction off crystal planes 
tilted by an angle $\alpha$ from the normal vector $\vec{n}$ of the crystal 
if the scattering angle $\theta_{\rm B}$ meets the Bragg condition.
  Right: Concentric crystal rings with different radii $r_1$ and $r_2$
and different tilted angles diffract X-rays coming from the right side, 
if their energies $E_1$ and $E_2$ satisfy the Bragg condition, onto
the focal spot on the left side of the figure (from \cite{2017SPIE10566E..03B} with kind permission from the authors).
  \label{f:laue}}
\end{figure}
Laue lenses use rings of crystals to Bragg diffract the $\gamma$-rays from
cosmic sources onto the detector, as shown in Fig.\,\ref{f:laue} \cite{2005ExA....20..171H,barr,2014RScI...85h1101S,2017SPIE10566E..03B}.
The focal length $F$ is given by $F=r/\tan{(2\theta)}$,
where $r$ is the radius of a given ring of crystals, and $\theta$ is the Bragg angle of the crystals \cite{vonBallmoos_2005,2014RScI...85h1101S}.
The wavelength bandpass of a single crystal 
resembles   a delta function centered on the wavelength
where the Bragg condition is met. 
Broader bandpass lenses can be constructed by 
(i) using different rings of crystals with different crystal orientations,
(ii) employing crystals with a mosaic structure and thus with an angular spread of the diffracting planes, or 
(iii) implementing curved diffracting planes.

The CLAIRE mission used a 45\,cm diameter, 2.77\,m focal length Laue lens 
made of 556 Ge-Si crystals to achieve an angular resolution of 
18 arcmin at 170\,keV \cite{2005ExA....20..171H}. 
Small amounts of Si were added to the Ge during the crystal growth process 
to make mosaic crystals with a bandpass of 3\,keV centered on 170 keV.
The experiment was flown on a stratospheric balloon flight in 2001 
and successfully detected $\gamma$-rays from the Crab Pulsar and Nebula \cite{vonBallmoos_2005}. 
The team is evaluating the merits of lenses made of Ge, Cu, SiGe, and 
commercially-produced mosaic gold crystals in order to reach higher energies.
The test samples achieve the theoretical maximum 
efficiency of 31\% at 600 keV \cite{barr,2017SPIE10566E..03B}.
The focal length of a Laue lens scales with the $\gamma$-ray energy. CLAIRE's 2.77\,m 
focal length at 170\,keV thus corresponds to a 511 keV focal length of 8.3\,m.
This focal length is approximately the same as that of the balloon-borne 
{\it X-Calibur} experiment (focal length: 8\,m, flight in 2018/2019) \cite{kislat_etal_2017}, and only 67\% of that 
of the upcoming {\it XL-Calibur} experiment (focal length: 12\,m, flights in 2022/2024) \cite{2021APh...12602529A}.
Operating Laue lenses on balloon-borne payloads is thus demonstrably possible.

Smither reviews the Laue lens developments over the last few decades and presents 
experimental and theoretical results for lenses with focal lengths from a few meters to a few hundred meters \cite{2014RScI...85h1101S}. A recent breakthrough in lens fabrication 
was achieved by making a lens from larger bent crystals rather than from hundreds of 
small crystals that need to be individually mounted and aligned.
High detection efficiencies over a broad energy band can be achieved by separating the lens and detector by as much as 200\,m. The Toulouse collaboration envisions a 2.2\,m 
diameter lens with a 200\,m focal length. 
The concept uses bent and unbent crystals and is projected to achieve a detection 
area of $\sim$600\,cm$^2$ in the two bands from 460 keV to 522 keV and 
from 825 keV to 910 keV \cite{2006NIMPA.567..333B,2014RScI...85h1101S}.
\begin{figure}[t!]
  \centering
  \includegraphics[width=0.8\textwidth]{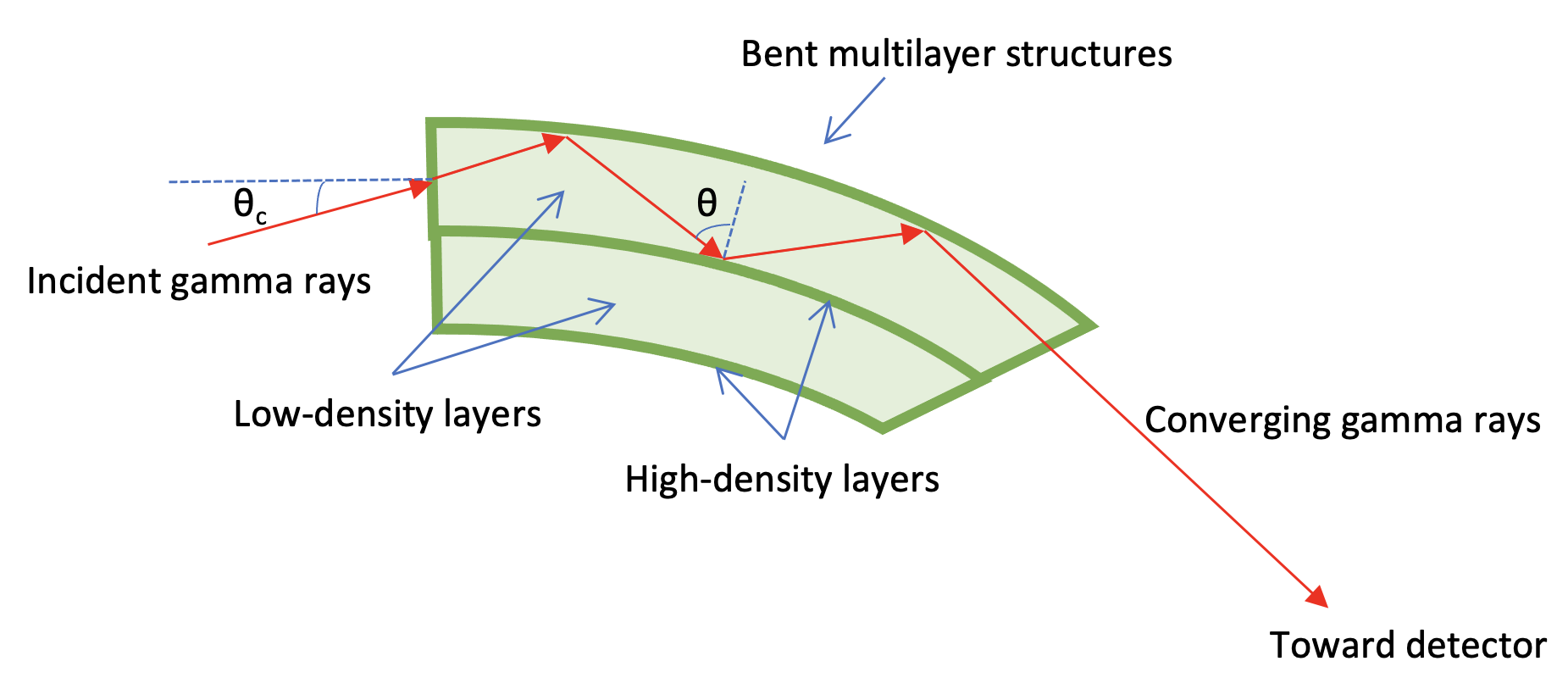}\quad%
  \caption{Diagram of channeling optics. The bent thin-film multilayer structure of alternating low- and high-density materials channel 
high-energy photons to a detector of smaller area. Photons arriving with an angle smaller than
a certain critical angle $\theta_{\rm c}$ are successfully channeled if the scattering angles $\theta$ 
inside the multilayer structure are within the correct bounds for total internal reflection to occur.}
  \label{f:concentrator}
\end{figure}

\begin{figure}[th]
  \centering
  \includegraphics[width=\textwidth]{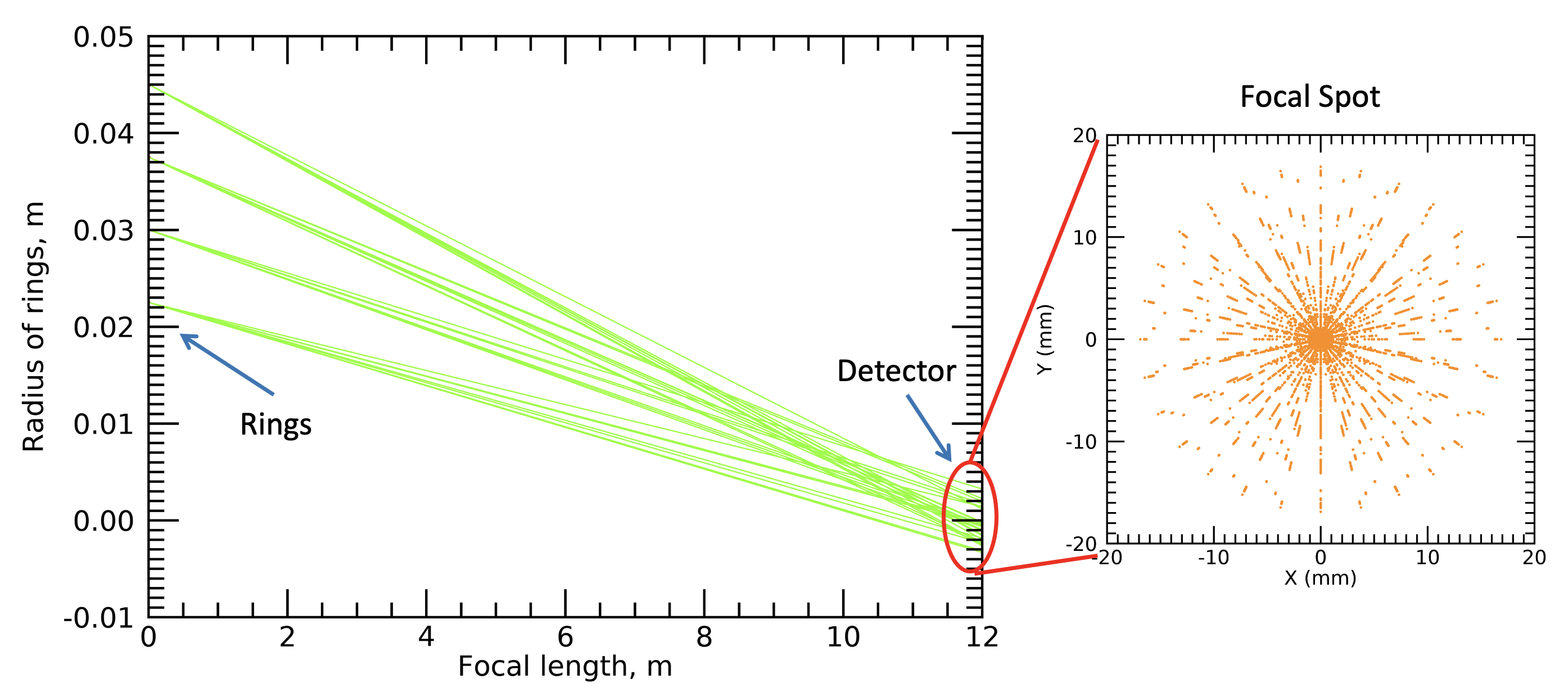}\quad%
  \caption{Ray-tracing output for parallel beams of 511\,keV $\gamma$-rays,
showing the concentrator rings, the intersection of the beams with the 
focal plane at 12\,m from the channeling optics, and the focal plane 
distribution of the $\gamma$ rays from a point source.}
  \label{f:tracing}
\end{figure}
The focal spots of the 2.77\,m focal length CLAIRE mission and the 200\,m focal length mission with an angular resolution of 15\,arcseconds
both have a diameter of 1.5\,cm.
A {\it 511-CAM} detector array with a sensitive area of $\sim$\,3\,cm diameter and 
$\sim$\,$1.3$\,$-$\,$1.45$\,mm diameter pixels would thus be a good match for either optical setup.
A 12\,m focal length lens with a 
$\sim$\,1\,arcmin angular resolution 
would produce a 3.5\,mm diameter focal spot, which would be nicely oversampled by
$\sim1.3-1.45$\,mm diameter pixels.

Finally, with regard to fabrication, we note that Wade et al.\ have developed techniques for assembling Laue lenses with excellent results \cite{2018NIMPA.895..135W} (see also \cite{Girou_2021}).

\subsection{Channeling Optics} 

As an alternative to Laue lenses, channeling optics, first proposed by Kumakhov et al.\cite{kumakhov_1990}, have recently received renewed attention \cite{1994SSRv...69..139R,2013SPIE.8861E..16D,2014SPIE.9144E..19B, 2018PhyU...61..980L,2018SPIE10699E..5VS,2015OExpr..23.8749K,2017SPIE10399E..1AS,2018ITNS...65..758T,2018PhDT.......215S}. Channeling optics use total external reflection off of gently curved multilayer structures, consisting of alternating layers of a high-density reflecting material and a low-density spacer material (Fig.\,\ref{f:concentrator}). This technique would allow $\gamma$-rays to be concentrated to a small, shielded detector in the focal plane \cite{bloser_etal_2015}. Since the photons can reflect multiple times within the structure, much larger bending angles can be achieved than for mirrors or Laue diffraction optics, resulting in much shorter focal lengths needed. Shirazi et al. \cite{2020JATIS...6b4001S} examined the 20\,cm diameter lens constructed with Ir/Si that produced a $\sim$\,1\,cm diameter focal spot over a $<$\,10\,m focal length and a 50-430\,keV energy range. The channeling process is effective over a broad energy band, as long as the angle of incidence is less than a certain critical angle and the photons are not excessively absorbed \cite{2018PhDT.......215S}. Smaller reflection angles and thus larger focal lengths can be used to extend the energy range to 511\,keV.

For this study, we re-created the $\gamma$-ray tracing code of the concentrator for a 12\,m focal length. The lens could be made of four concentric rings of channeling segments, with ring radii of 2.25, 3, 3.75, 
and 4.5\,cm and lengths of 2.1, 2.7, 3.5, 
and 4.6\,cm, respectively \cite{2017SPIE10399E..1AS}. Each ring would use 1\,cm-wide, 7.5\,mm-thick segments. The channeling multilayers are  
made of alternating magnetron-sputtered layers of  30 nm W and 150 nm Si 
and have total thickness of 5.4 $\mu$m each.
The channeling segments would have a 
total mass of 0.5 kg.

Altogether, the concentrator would produce a $\sim$\,3.5\,cm diameter focal spot. The channeling optics would have a field of view with an angular radius of 4.47\,arcmin  (Fig.\,\ref{f:tracing}).
Shirazi et al. discuss Cadmium Zinc Telluride (CZT) and Si detectors with a few cm$^2$ of detection area as focal plane instrumentation \cite{2018ITNS...65..758T}. The {\it 511-CAM} setup would be an excellent fit with such channeling optics, improving greatly upon the energy resolution of the CZT and Si detectors.

\section{The Focal Plane Instrumentation for the {\it 511-CAM}}
\label{s:511-Cam}
\subsection{Microcalorimetric \texorpdfstring{$\gamma$}{Gamma}-Ray Detectors}
\begin{figure}
  \centering
  \includegraphics[width=.3\textwidth]{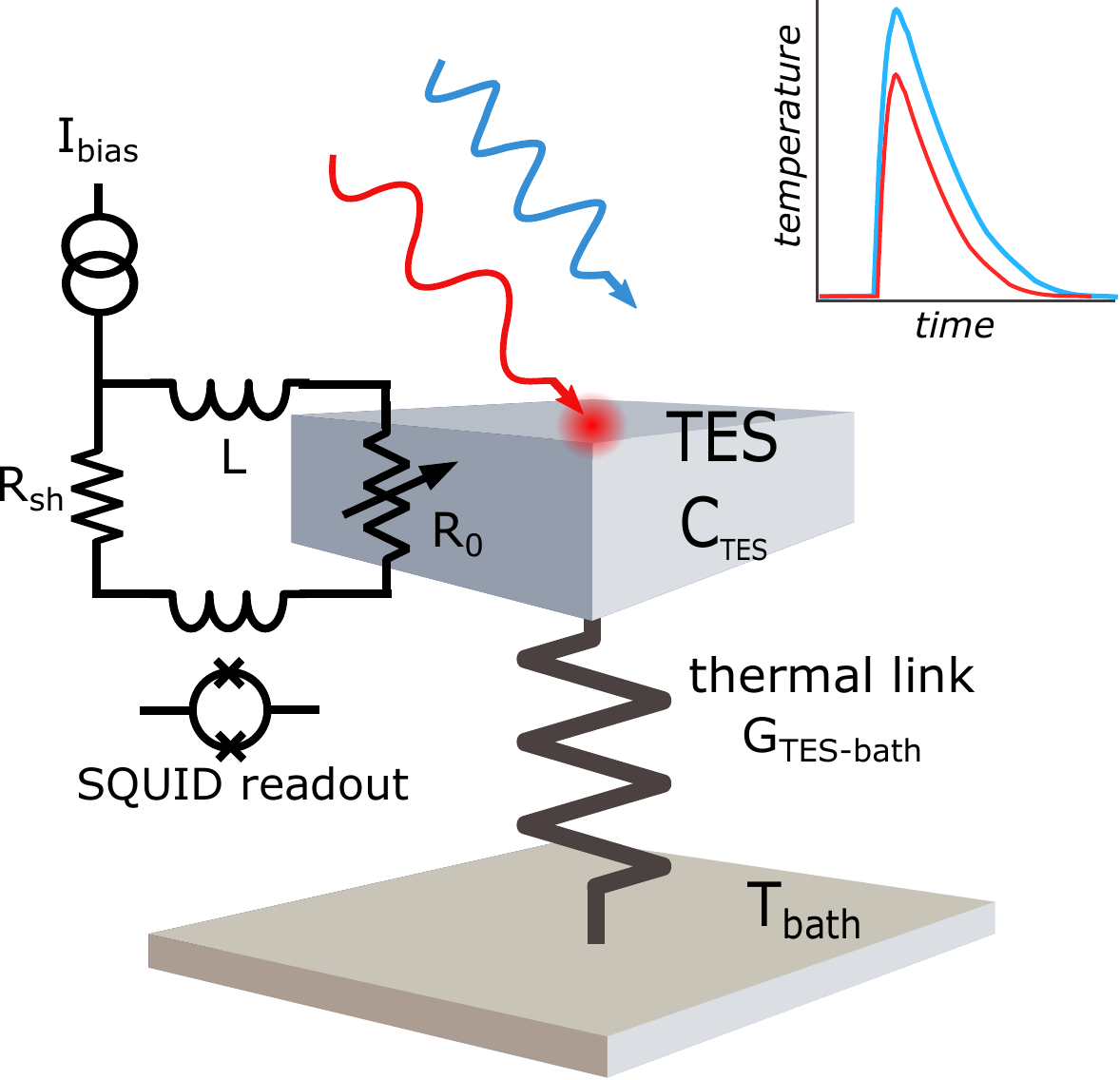}\hspace*{1.5cm}
  \includegraphics[width=.3\textwidth]{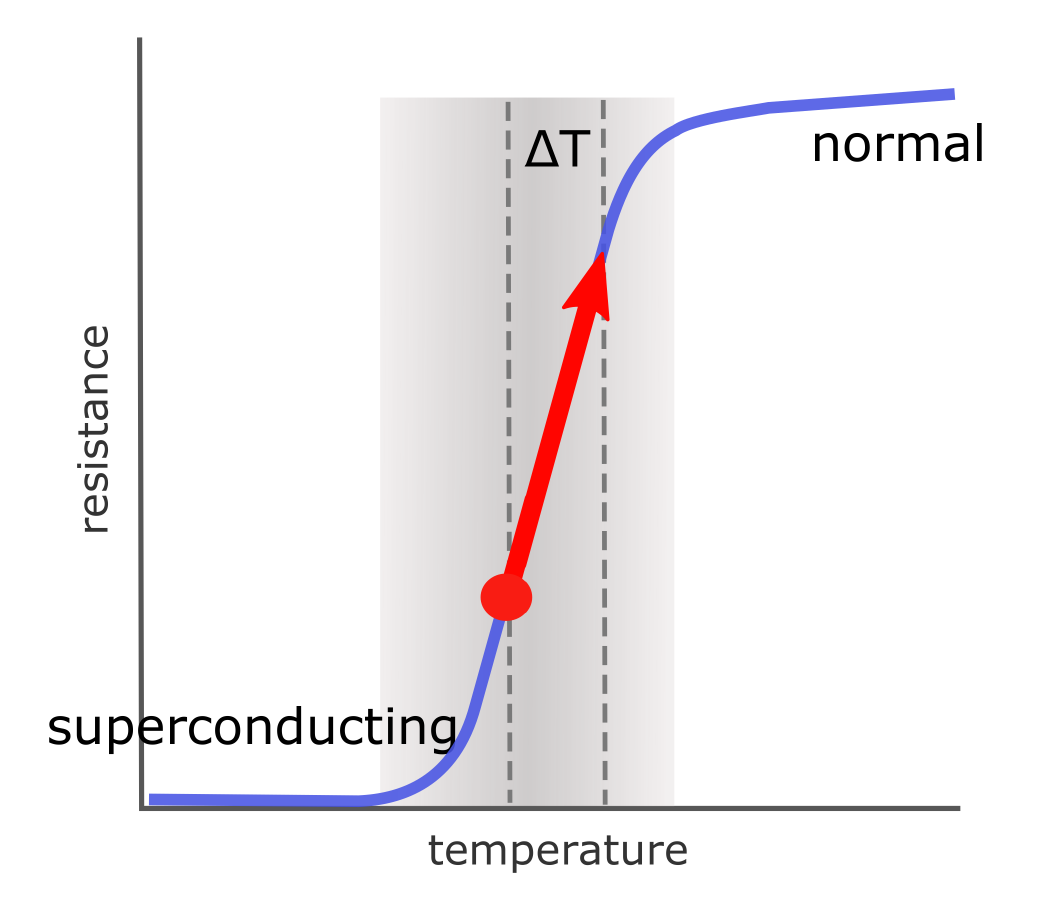}
  \caption{Microcalorimeters use an absorber thermally coupled
  to a Transition Edge Sensor (TES), shown in the left panel 
  as a single unit. The temperature transient due to an incident 
  $\gamma$-ray causes a large change of the bias current through 
  the TES, operated at the normal-superconducting transition, as shown in the right panel.}
  \label{fig:micro}
\end{figure}
A microcalorimeter detector operates at $\sim$\,100\,mK temperatures and records 
$\gamma$-rays via the measurement of the temperature transients that the $\gamma$-rays 
cause when they strike the detector absorbers (see Fig.\,\ref{fig:micro}, left panel). 
Each absorber is thermally coupled to a Transition Edge Sensor (TES), which is a thin superconducting film that works as a thermometer.
The TES is biased at the lower end of its superconducting-to-normal transition.
Owing to the sharpness of the transition, a tiny temperature change translates into 
a large change of the current allowed through the TES (Fig.\,\ref{fig:micro}, right panel).

\begin{figure}
\centering
\includegraphics[width=.5\textwidth]{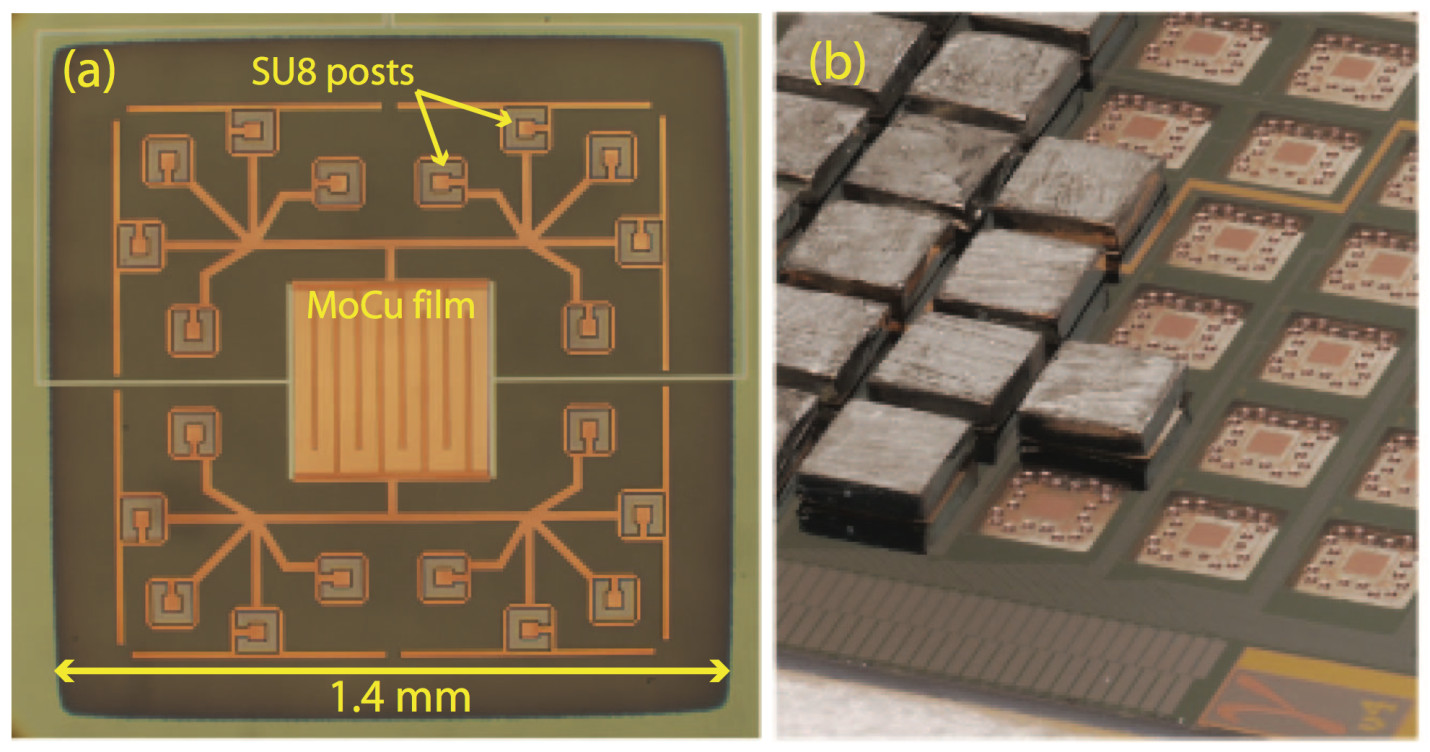}\hspace*{0.2cm}
\includegraphics[width=.3\textwidth]{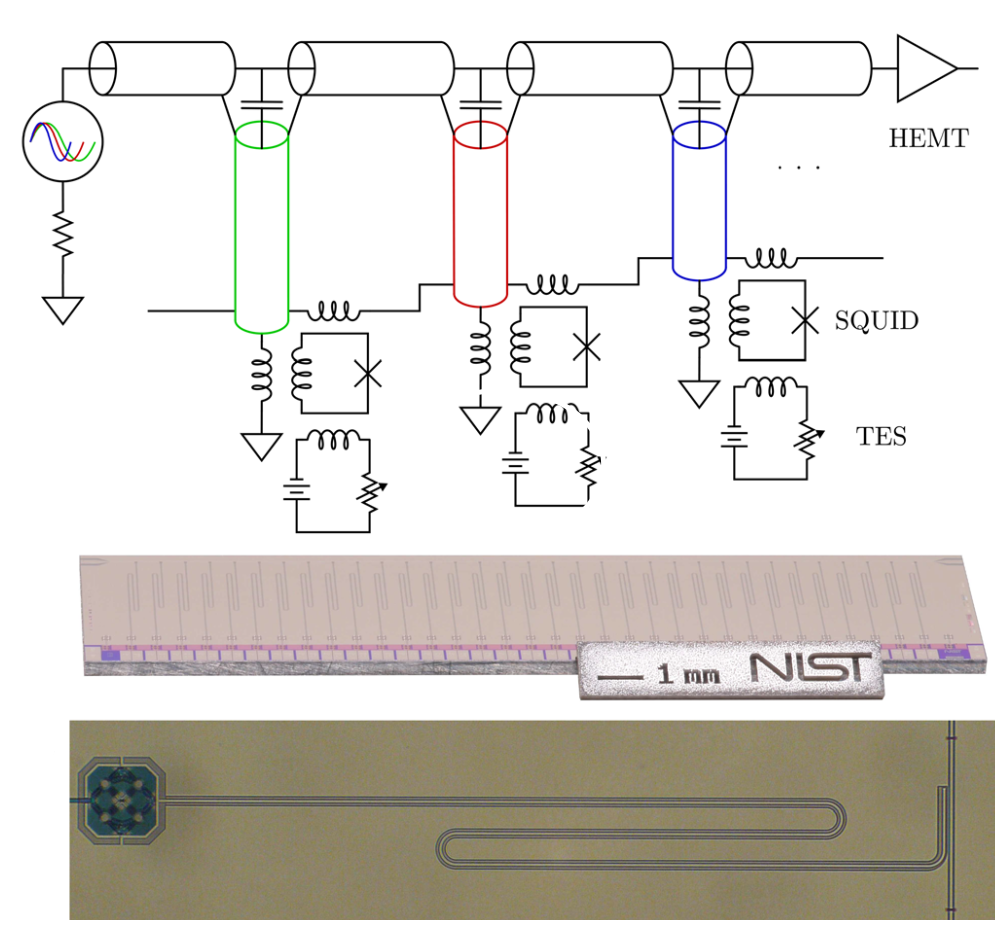}
\caption{Left: Image of a \sh Mo-Cu TES thermally 
coupled to SU8 epoxy posts via Cu traces. Center: Image showing
the \sh base plate after partial installation of
1.45$\times$1.45$\times$0.38~mm$^3$ Sn absorbers glued onto the epoxy posts which are mounted on the TES sensors. 
Right: The top right panel shows the multiplexing scheme: the pixels (not shown here) are connected to transition edge sensors (shown here right above the two photographs 
of resonators), and the TES signals are sensed with
SQUIDs, which modulate the resonant frequencies of resonators. 
The resonators, shown in color, are read out with a common microwave feedline.
The two photographs in the bottom right show 33 resonators 
(center right panel) and a single resonator (lower right panel).}
  \label{fig:tes-image}
\end{figure}
\begin{figure}[tb]
  \centering
\includegraphics[width=.5\textwidth]{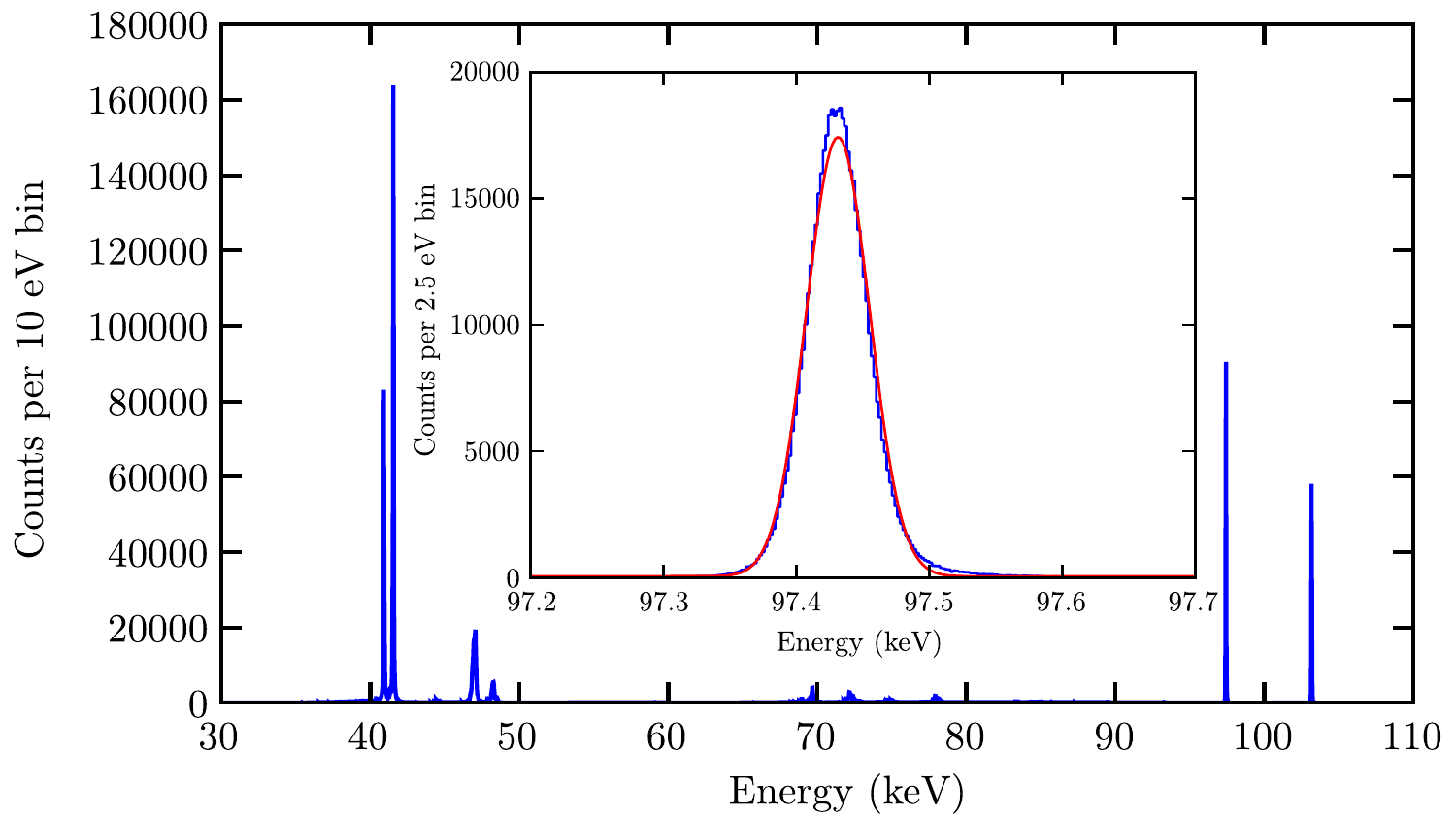}
  \caption{
   $^{153}$Gd energy spectrum obtained with a \sh detector with $1.45\times1.45\times0.38$\,mm$^3$ pixels.
   The inset shows the 97\,keV $\gamma$-ray line with a FWHM of 55\,eV \cite{mates_etal_2017}.}
  \label{fig:spec}
\end{figure}

  
The NIST group has developed large microcalorimetric $\gamma$-ray 
detector arrays that are well-suited for NASA's balloon-borne and space-borne 
astrophysics missions. The \sh detectors and their successors use $\sim$\,cubic-millimeter-sized absorbers attached to TES sensors, and feature
a fully scalable microwave readout scheme. Figure \ref{fig:tes-image} 
shows the key components of the \sh detectors:
\begin{enumerate}
\item The proximity effect is used to make {MoCu} (or MoAu) transition edge sensors with transition temperatures $T_{\rm c}$ tailored to 
$\sim$100\,mK (left panel).
\item Sn absorbers are mounted on SU8 epoxy pillars that mechanically 
support the absorbers and establish the thermal link between 
the absorbers and the TES sensors (center panel).
\item Each TES is read out by a Superconducting Quantum Interference 
Device (SQUID) that is AC-coupled to a GHz microwave resonator (right panel).
\end{enumerate}
A $\sim$\,250-500 pixel detector box can then be read out with a single microwave feedline interrogating the resonators of all the pixels.
The SQUIDs are operated with a modulated magnetic field flux bias, 
enlarging the dynamic range that can be satisfactorily sampled with 
the SQUIDs. The microwave multiplexing is ideally suited for 
astronomical applications, where many pixels are needed and 
complexity and mass need to be minimized while achieving 
a high degree of ruggedness.
The \sh detectors, used with 1.45$\times$1.45$\times$0.38\,mm Sn pixels, achieve 
an energy resolution of 55\,eV FWHM over much of their dynamic range (Fig.\,\ref{fig:spec}) \cite{2012RScI...83i3113B,noroozian_etal_2013,mates_etal_2017}.
\subsection{Optimization of the {\it 511-CAM} Focal Plane Instrumentation}
\label{s:focal}
Whereas different absorber materials have been tested extensively for X-ray micro-calorimeters, fewer studies have been completed for 
$\gamma$-ray microcalorimeters\cite{2015ITAS...2580173I,2017IEITE.100..283O}.
Given that the performance of $\gamma$-ray microcalorimeters is limited by phonon noise
that scales with the heat capacitance of the absorbers, the selection of an absorber 
material with low specific heat capacity is particularly important 
for these $\gamma$-ray detectors.

We consider four materials as options for the {\it 511-CAM} absorbers: 
Sn (the material currently used on the \sh detectors), Ta, Bi, or a Bi$_{0.57}$-Sn$_{0.43}$ alloy (see Table \ref{t:pixel}).
\begin{table}[H]
    \centering
    \begin{tabular}{|p{1cm}|c|p{1.3cm}|p{1.7cm}|p{1.7cm}|p{1.7cm}|p{1.3cm}|p{3cm}|}
    \hline
    Mate-rial & $<\!\!Z\!\!>$ & density [g/cm$^{3}$] & $\alpha_{\rm pe}$ [cm$^2$/g] & $\alpha_{\rm sc}$ [cm$^2$/g] &
    $C_{0.1K}$ [J/K/m$^{3}$]  & {\small Refs} & Comments\\ \hline \hline
    Sn & 50 & 7.31 & 1.64$\times10^{-2}$ & 7.73$\times10^{-2}$ &  1.3$\times10^{-2}$ & \cite{2008JLTP..151..413B, 2012RScI...83i3113B,noroozian_etal_2013,2015ITAS...2580173I,mates_etal_2017,2017IEITE.100..283O} & Currently used for \sh  \\  \hline
    Bi & 83 & 9.78 & 8.62$\times10^{-2}$ & 7.93$\times10^{-2}$ & 3.7$\times10^{-2}$ & \cite{2017ApPhL.111s2602Y,2018JLTP..193..225Y,2008JLTP..151..413B,2020JLTP..199..393H} & {\it {\small 511}{\scriptsize-CAM}}{\small \,reference\,design}\\ \hline
    Ta & 72 & 16.6 & 5.71$\times10^{-2}$ & 7.81$\times10^{-2}$ &  1.0$\times10^{-2}$ & \cite{2015ITAS...2580173I,2017IEITE.100..283O} & {\small Potential issue:\newline H-contamination, which may increase the heat capacity measured} \\ \hline
    BiSn & 69 & 8.6 & 5.62$\times10^{-2}$ & 7.85$\times10^{-2}$ & tbd & none & {\small \,57\%\,Bi,\,43\%\,Sn}\\  \hline 

    \end{tabular}
    \caption{Properties of promising 511\,keV absorber materials. 
    The references point to measurements or estimates of the 100\,mK
    heat capacitances and the performance of the material as a
    microcalorimeter absorber. Here, $\alpha_{\rm pe}$ and $\alpha_{\rm sc}$ are the photoelectric effect and scattering absorption coefficients at 511\,keV from NIST's XCOM database.\cite{NIST:XCOM} \label{t:pixel}}
\end{table}

\begin{figure}[t]
  \centering
  \includegraphics[width=.62\textwidth]{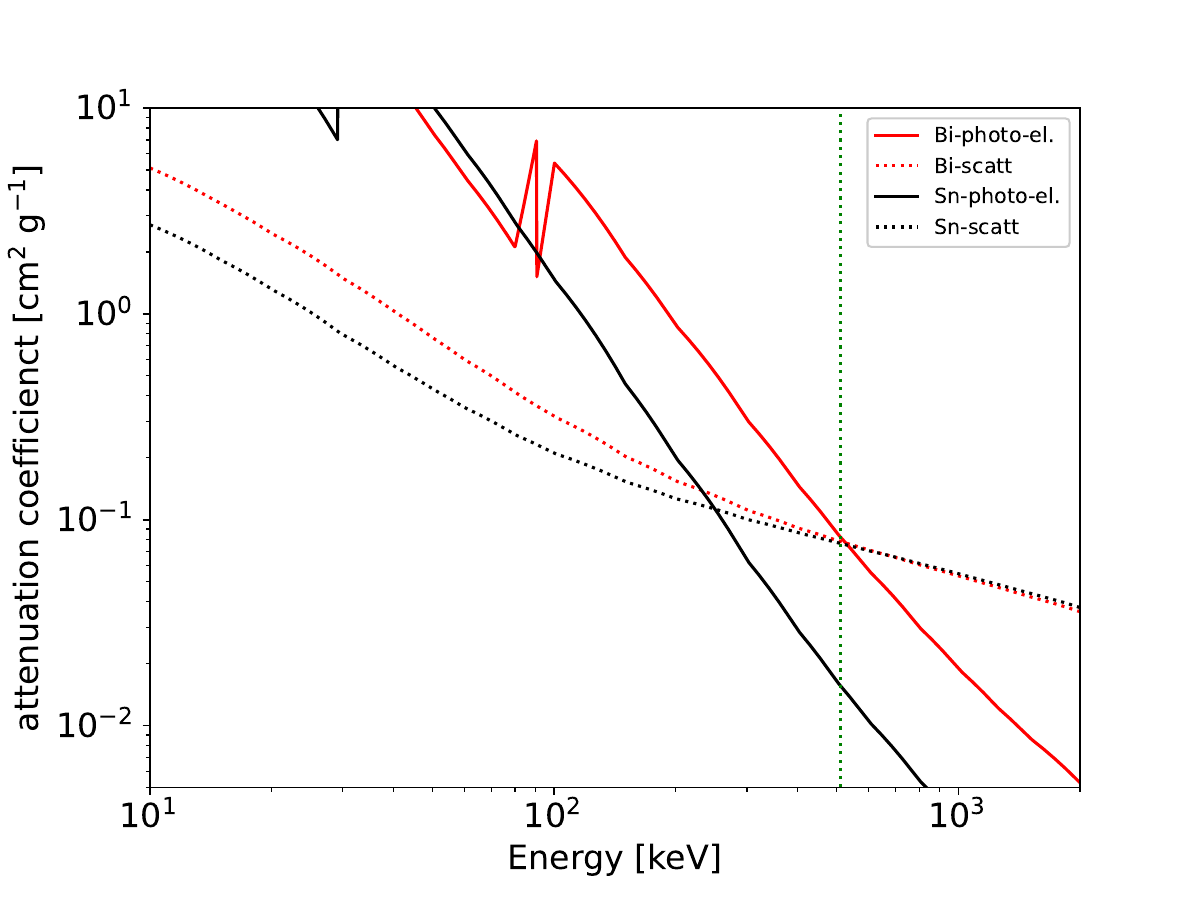}
  \caption{Comparison of the photoelectric effect (solid lines) and scattering (coherent and incoherent, dotted lines) attenuation coefficients for
Bi (red lines) and Sn (black lines) as a function of photon energy.  Note that the cross-sections due to pair production gain prominence at slightly higher photon higher energies than shown here.}
  \label{f:BiSn}
\end{figure}

Figure \ref{f:BiSn} compares the photoelectric effect and scattering cross-sections of 
Sn and Bi in the 10 keV to 2 MeV energy range. 
Whereas the 511 keV Compton scattering cross-sections are similar for both materials, 
the photoelectric cross-section of Bi is 5.3 times larger than that of Sn. 
Photoelectric effect events, compared to Compton scattering events, are particularly valuable, since they lead to a smaller number of interactions and thus yield the better energy resolution. In this regard, Bi is a superior choice to Sn. 

Another key property of an absorber material is the heat capacity. 
The fundamental limit on the energy resolution of a microcalorimetric detector is given by
\begin{equation}
\Delta E\,=\,2.35 \times \sqrt{4 k_{\rm B} T C (1/\alpha) \sqrt{n/2}}.
\label{e:irwin}
\end{equation}
Here, $k_{\rm B}$ is the Boltzmann constant, $T$ is the temperature of the absorber, 
and $C$ is the heat capacitance of the absorber and TES.
The parameter $\alpha\,=\,(T/R)(dR/dT)$ characterizes the steepness 
of the superconductivity-to-normal transition. The parameter $n\approx 4.6$ 
depends on the heat transport between the TES electrons 
and the thermal bath \cite{1995ApPhL..66.1998I}. 
Bi beats Sn as an absorber material for the {\it 511-CAM} in this heat capacity regard as well, given that the product of the
volumetric heat capacitance times the photoelectric effect absorption length is 
4.4$\times 10^{-4}$\,J\,K$^{-1}$\,m$^{-2}$ and 1.1$\times 10^{-3}$\,J\,K$^{-1}$\,m$^{-2}$ for Bi and Sn, respectively
(see \cite{2008JLTP..151..413B}  for the heat capacitance of Bi).
As seen in the formula above, the lower the heat capacitance, the better the energy resolution. Bi is furthermore preferred over Si as it offers a $\sim$5 times higher ratio of
photoelectric effect to Compton scattering cross section, leading to a smaller
number of interactions, and thus to better energy resolutions.

In terms of fabrication, we will use Bi cubes made from high-purity Bi sheets
\cite{2017ApPhL.111s2602Y,2018JLTP..193..225Y,2020JLTP..199..393H}. The Bi sheets can be obtained from companies such as those listed in the references \cite{goodfellow, edgetech, vulcan}.

Possible alternatives to Bi include Ta and Bi$_{0.57}$Sn$_{0.43}$. 
Like Bi, Ta exhibits a lower heat capacitance per area than Sn and achieves excellent 
resolutions at X-ray energies.  
{For example, a microcalorimeter with small Ta absorbers of dimensions 0.5\,mm\,$\times$\,0.5\,mm\,$\times$\,0.3\,mm has been shown to achieve a 445 eV FWHM energy resolution at 662 keV, thus solidly outperforming HPGe detectors. Even though this study measured the capacity of Ta to be higher than theoretically-calculated, it also notes that H-adsorption by Ta may have led to this higher heat capacity. This factor, along with vibrational noise from the dilution refrigerator, potentially contributes to the slightly worse-than-expected energy resolution.} \cite{2017IEITE.100..283O}.
The motivation for using Bi$_{0.57}$Sn$_{0.43}$ is to make an 
absorber material that combines the high stopping power of Bi with the 
low heat capacitance of a superconductor that transitions at a higher, easier temperature. While the transition temperature of 
Sn at ambient pressures is 3.72\,K, that of Bi is estimated to be 
$<$1.3\,mK \cite{bismuth, bismuth2}; Prakash et al.\ reported 0.53\,mK \cite{2017Sci...355...52P}. 
The transition temperature $T_{\rm c}$ hardly changes from 3.7 K 
across the alloy series until the alloy contains 90\% Bi or more \cite{1968PhLA...27..116G}. Thus, one could include substantial amounts of Bi in the alloy (with its higher photoelectric cross-sections) without significantly budging the more ideal, higher superconducting transition temperature. 

However, yet another crucial figure of merit is the energy resolution per triggered pixel. Here, we provide an initial estimate of the projected energy resolution {achievable if we focus on the possibility of using Bi as the absorber material}. The {\it 511-CAM} will adopt the second-generation absorber/TES layout with $1.3\times1.3$\,mm$^2$ footprint absorbers. Scaling the 55\,eV FWHM energy resolution of the \sh detectors with the square root of the heat capacitance of the absorbers (Equation \ref{e:irwin}), we estimate that the {\it 511-CAM} would achieve a per-pixel energy resolution of 230\,eV FWHM. 
Given the heat capacity of the Bi absorbers, we expect that 511 keV photons will cause
a similar temperature excursion in the Bi absorbers as 30 keV photons do in the current absorbers. Saturation caused by the TES being heated to normal conductivity will thus not be an issue.
This energy resolution improves dramatically over the best theoretically possible (Fano factor limited) 511 keV energy resolution of $\sim$900\,eV FWHM of HPGe detectors. 

\section{Case study: a balloon-borne 12\,m focal length {\it 511-CAM} mission} 
\label{s:telescope}
The {\it 511-CAM} technology enables missions with different $\gamma$-ray optics, 
focal lengths, and field of views. 
In this section, we discuss an exemplary balloon-borne mission. Long Duration Balloon (LDB) flights and Ultralong Duration Balloon Flights (ULDB) can be as long as 55 days and 100 days, respectively. The LDB flights typically reach altitudes of 38 km.

\begin{figure}[t!]
  \centering
  \includegraphics[width=0.52\textwidth]{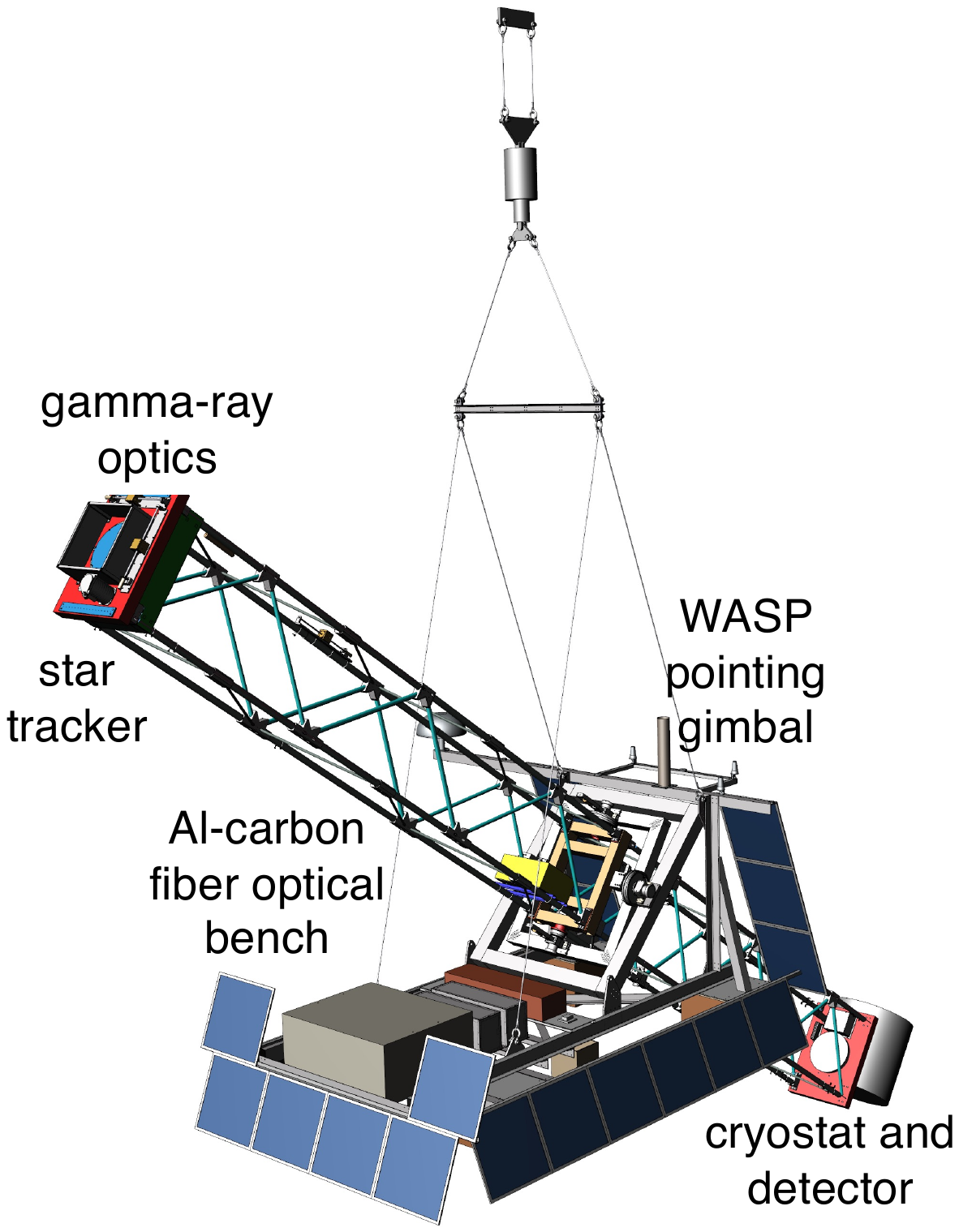}\hspace*{0.25cm}
  \includegraphics[width=0.47\textwidth]{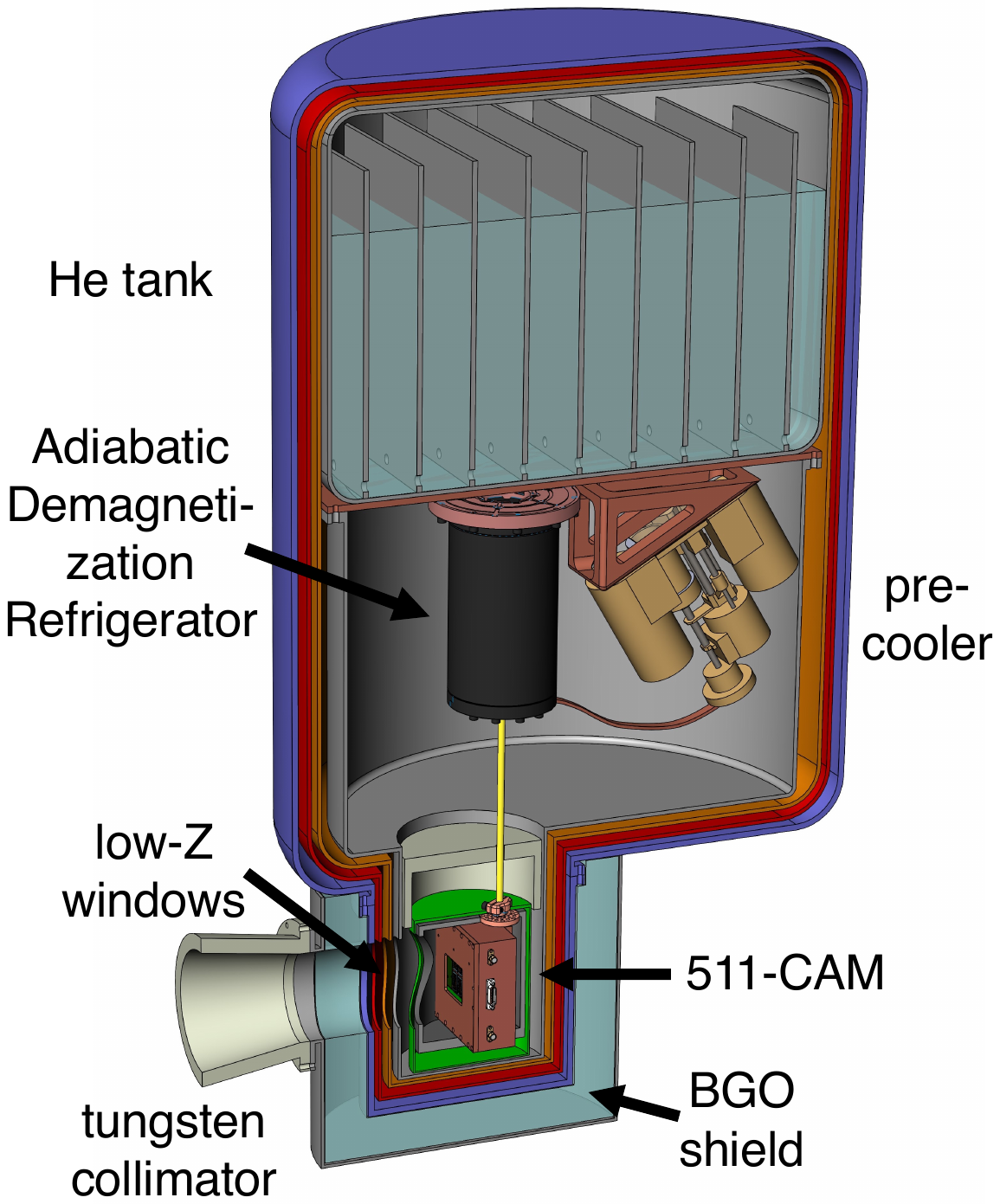}
  \caption{Sketch of a 12\,m focal length {\it 511-CAM} enabled mission.
  Left: Pointed 12\,m focal length balloon payload with $\gamma-$ray concentrator at the front end and the cryostat at the rear end. Right: Cryostat design with an Adiabatic {Demagnetization} Refrigerator getting to 100 mK, after a pre-cooler first achieves 300\,mK. LHe is used as the cryogen. The Nb and cryoperm (A4K) shields are shown in green.}
  \label{f:mission}
\end{figure}

The design uses a 12\,m focal length channeling concentrator (Fig.\,\ref{f:mission}). The {\it 511-CAM} mission inherits {\it XL-Calibur}'s 12\,m 
optical bench design, which uses carbon fiber tubes, Al joints, and Al honeycombs \cite{2021APh...12602529A}.
This setup draws from the designs of the {\it XL-Calibur} hard
X-ray polarimetry mission scheduled for launches in summer 2022 and 2024, 
as well as the {\it ASCENT} mission proposed to the 
2020 Pioneers opportunity \cite{ascent,2023arXiv230101525K}.
{\it ASCENT} employs a single- but larger-area 
\sh-type focal plane detector to map the
$^{44}$Ti hard X-ray lines from supernova remnants such as Cas-A.

The {\it XL-Calibur} optical bench has a mass of $\sim$\,300 kg, and is designed for a $>$\,16 factor of safety to guarantee that it survives the parachute deployment and landing after the balloon flight. The telescope is pointed with the Wallops Arc-Second Pointer {\it WASP}, achieving a pointing stability of better than 1 arcsec \cite{2014cosp...40E3221S}. 
The {\it XL-Calibur} science payload includes a 65 kg X-ray mirror and a 70 kg detector (made of Cadmium Zinc Telluride pixels), active shield (BGO), and readout (CPUs, various electronics) assembly.
A {\it 511-CAM} mission would require somewhat heavier components, i.e.,\ 
a $\sim$\,250\,kg cryostat, detector, and 
shield assembly instead of the 70 kg one used for
{\it XL-Calibur}.

The detector is operated inside a wet LHe cryostat. 
We are considering two refrigerator options for cooling the detectors 
to $\sim$85\,mK temperatures.
The first option uses an Adiabatic Demagnetization Refrigerator (ADR) \cite{2010AIPC.1218..633W}; the second option uses a mini-Dilution Refrigerator (mini-DR) \cite{chase_dilutor}.
Whereas Fig.\,\ref{f:mission} shows the ADR solution, we are currently building a 
prototype for the mini-DR solution.
The mini-DR achieves an excellent 10\,$\mu$K Root Mean Square (RMS) temperature stability,
even when the unit experiences continuous tilting (as required for a unit in the focal plane
of a balloon borne telescope) \cite{ryan}. The temperature stability is comparable to 
the 4\,$\mu$K stability routinely achieved with ADRs \cite{2019RScI...90l3107S}.
For details of the mini-DR control, see Table 1 of \cite{ryan}.
Independent of the refrigerator solution, the only cables going to just one {\it 511-CAM} 
detector array are four coax cables (two ``in'' and two ``out'', sufficient 
for the readout of all 8$\times$22$\times$22\,$=$\,3,872 pixels), 
and two twisted-pair cables, 
one for the TES biases and one for the SQUID flux ramps.
The expected heat dissipation of the {\it 511-CAM} of 
$\sim$\,1.2\,$\mu$W is dominated by the shunt resistors 
providing the TES biases. 
This power is below the cooling power provided by the pre-cooled ADR and the mini-DR.

The detector is shielded against cosmic rays and their
daughter products with a  passive tungsten shield at the inside of the 
cryostat and an active BGO shield at the outside.
Magnetic shielding is provided by a superconducting 
niobium shield and a 0.062-inch thick Amumetal 4K shield.
The \sh SQUIDs are designed to be highly gradiometric, making them nominally insensitive to uniform magnetic fields. 
In practice, this means that a SQUID with a footprint of 40,000\,$\mu$m$^2$ shows an effective pickup area of only
$1-4$\,$\mu$m$^2$. We estimate that the magnetic fields
do not affect the energy resolution of the {\it 511-CAM}, even during re-pointings of the instrument.  
{The detector will see the gamma-ray sky through a 
suitable window. Whereas a thorough optimization still needs to be performed, we note that} a possible solution could be a Light Element X-ray High Transmission (LEX-HT) window from Luxel.
The window can be mounted on an ISO 100 flange with a 25.4\,mm aperture, to let the $\gamma$-rays in. The filter is made of a 110\,nm Al / 200\,nm LUXFilm Polyimide film, secured with a stainless-steel ring, to block photons coming from a variety of temperatures from reaching the inside of the cryostat \cite{luxel}. Since Al is a broadband IR-blocking material, it does not require precise tuning of its geometry or thickness to block the blackbody radiation emitted by any given temperature. { Thicker windows-filter combinations may still give suitable $>$10\,keV transmissivities while being mechanically more rugged.} 

The detector gain (i.e., the relation between the true photon energy and the detected signal)
may drift with time owing to thermal contributions and amplifier 
contributions \cite{2016JLTP..184..498P}.
As mentioned above, we expect that we can keep the temperature of the cold head stable to within
$\sim$\,4-10\,$\mu$K RMS \cite{2019RScI...90l3107S,ryan}, and to 
correct for gain variations based on in-flight calibrations with a radioactive $^{133}$Ba source.
This source emits $\gamma$-rays at 80.9971\,keV (34.06\%), 276.398\,keV (7.164\%), 302.853\,keV (18.33\%), 356.017\,keV (62.05\%), and 383.851\,keV (8.94\%). 
On the ground, before launch, we will calibrate by examining how the lines from both the $^{133}$Ba source and a $^{22}$Na source (which emits 511\,keV $\gamma$-rays) change with respect to the gain, in order to know how the line changes at 511\,keV given a certain change in the $^{133}$Ba lines during flight \cite{2019RScI...90l3107S}.
Note that the $^{133}$Ba source contributes background mainly below its 383.851\,keV highest-energy $\gamma$-ray line, well below the 511-keV energy range of interest.

Detailed descriptions of the cold and warm detector readout electronics and 
the data analysis methods  can be found in \cite{2011PhDT........55M,mates_etal_2017,2018JLTP..193..485G,becker_2019}.
Whereas our current lab system uses the Reconfigurable Open Architecture Computing 
Hardware ROACH2 platform \cite{2018JLTP..193..485G} to interrogate the detectors with 
microwave tones and to digitize the detector output, we are currently transitioning 
to using Xilinx’s Zynq UltraScale$+$ Radio Frequency System-on-Chip (RFSoC) \cite{xilinx}. 
The RFSoC operates at GHz frequencies without the need to downsample and upsample the signals.

During Long Duration Balloon flights, data can be telemetered to the ground via
Tracking and Data Relay Satellite System (TDRSS) at typical rates of 800 kbit s$^{-1}$.
Scaling the cosmic ray and atmospheric background measured with the {\it XL-Calibur} 
mission \cite{2022arXiv221204139I} according to the {\it 511-CAM} and {\it XL-Calibur}
detector mass, we expect a $>$\,15\,keV background rate of 60 Hz. 
Choosing a calibration source that generates an equal rate of calibration events, 
and accounting for the fact that the signal rate is negligible, we predict a detection 
rate of 120\,Hz, with each event generating 4\,kBytes of data.  
We envision the storage of all the raw data on a solid state disk located on 
the balloon gondola. An on-board PC 104 computer will be used to perform 
on-board processing of the data and will send, for each event, 20 Bytes of summary 
information (pixel number, trigger time, and reconstructed event amplitude) to the ground. 
The telemetered data will be used for system health checks. 

We are currently preparing for the test of the mini-DR described in \cite{ryan} 
and a 36-pixel prototype TES array on a one-day stratospheric balloon flight 
from Fort Sumner (NM) in Fall 2023 or Fall 2024. A detailed description of the DR-TES 
(Dilution Refrigerator-Transition Edge Sensor) payload, including the cryostat 
and detector design, the control, data acquisition, and telemetry software, 
the warm readout electronics, and the detector calibration, will be given in forthcoming papers.
\section{Simulations of the focal plane detector\label{s:perf}}
\begin{figure}
  \centering
  \includegraphics[width=0.8\textwidth]{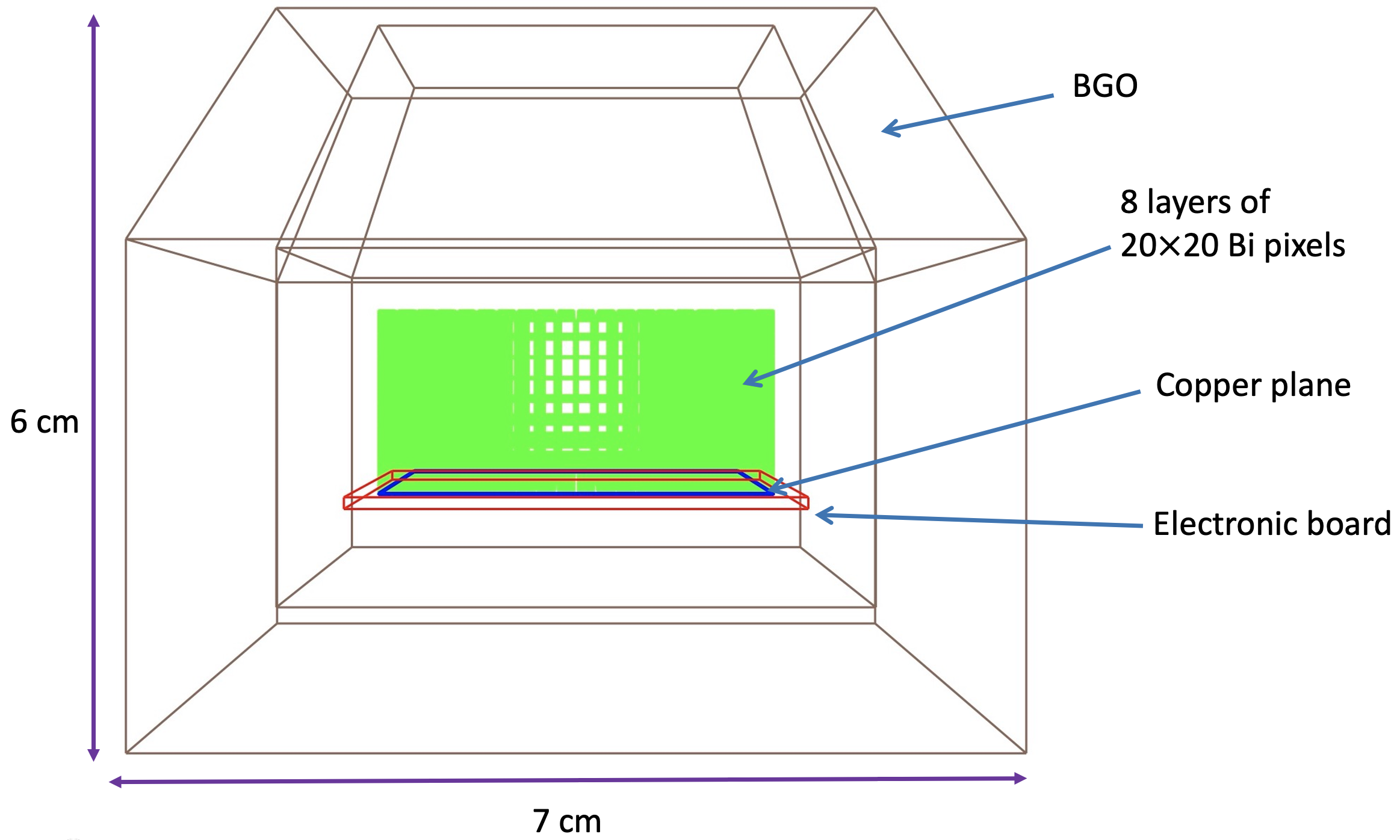}
  \caption{Schematic design of the {\it 511-CAM} mass model, with the shielding and some passive materials included.}
  \label{fig:Geometry}
\end{figure}

We simulated the performance of the {\it 511-CAM} technology described in this study, as a potential focal plane detector for the $\gamma$-ray concentrator, using the ``MEGAlib" software tools \cite{2006NewAR..50..629Z}. We created a realistic geometry and described an idealized mass model for a simple detector, neglecting much of the passive material between detector layers. Note that the passive materials (carrier boards, copper planes, copper traces, TES sensors, SQUIDs, and resonator circuitry) are either low-Z or are very thin ($<1\,\mu$m) and absorb much less energy than the 2\,mm thick Bi absorbers. We assumed a reference configuration that uses 8 detector planes, each carrying 20$\times$20 Bi pixels (1.45$\times$1.45$\times$2\,mm), which covers a $\sim$3.47$\times$3.47\,cm$^2$ focal plane area. 
We chose to make the {511-CAM} out of 8 layers of Bi.
This choice results in absorbing 93\% of the 
photons that hit the absorbers rather than the 
gaps between absorbers.
We assume a per-pixel energy threshold of 1.2\,keV. Although this trigger threshold 
is lower than the anticipated lower end of our dynamical range, the threshold 
barely affects the results, as most energy deposits will far exceed any likely threshold values. An active BGO shield surrounding the detector, 5\,cm thick at the bottom and 2\,cm thick on the sides, is included as the instrument shielding used to greatly reduce the background. An energy resolution of 10\,keV and a trigger threshold of 70\,keV is used for the BGO (Fig.\ \ref{fig:Geometry}). 
In the simulations, the BGO shield has a cylindrical hole, on the top, of the size of the focal spot that allows $\gamma$-rays to enter. The simulations neglect the Nb and cryoperm (A4K) magnetic shields for now.

Then, we utilized a 511\,keV near-field source whose profile is given by a concentrated radial beam onto the focal plane. An effective observation time of 1\,Ms has been assumed for a balloon flight, and the flux was modified for the effects of atmospheric absorption. The resulting output file contained the simulated interaction information, including the time, type, position, and energy deposition of each interaction. 

\begin{figure}
  \centering
  \includegraphics[width=.46\textwidth]{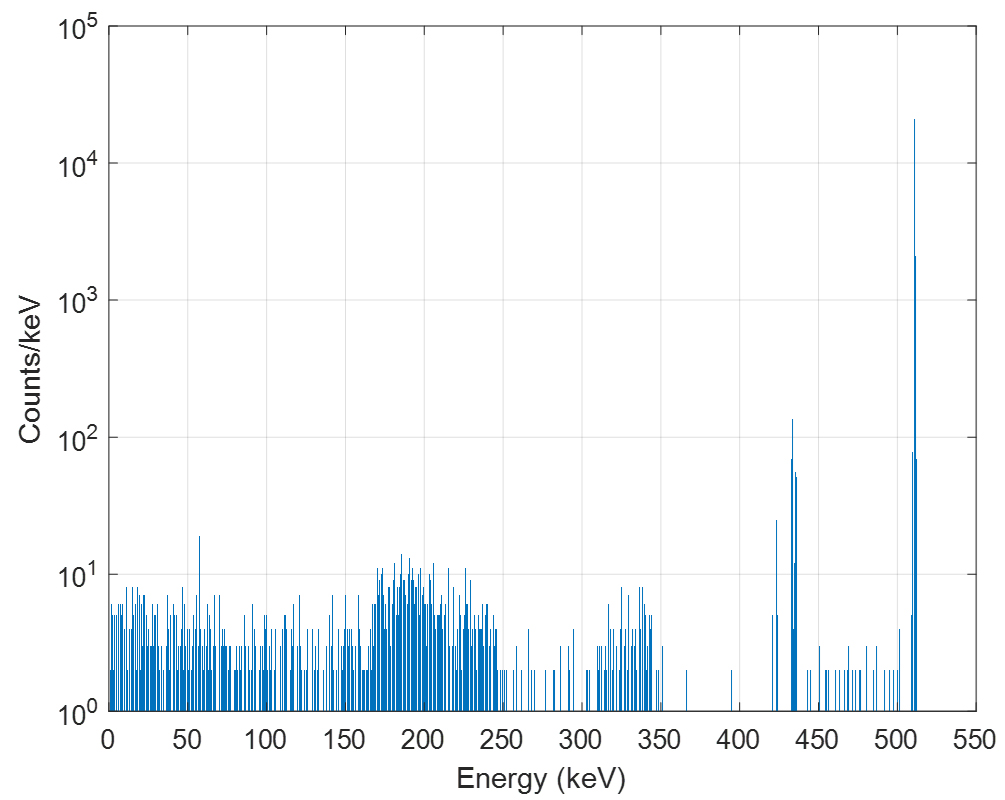}\quad%
  \includegraphics[width=.51\textwidth]{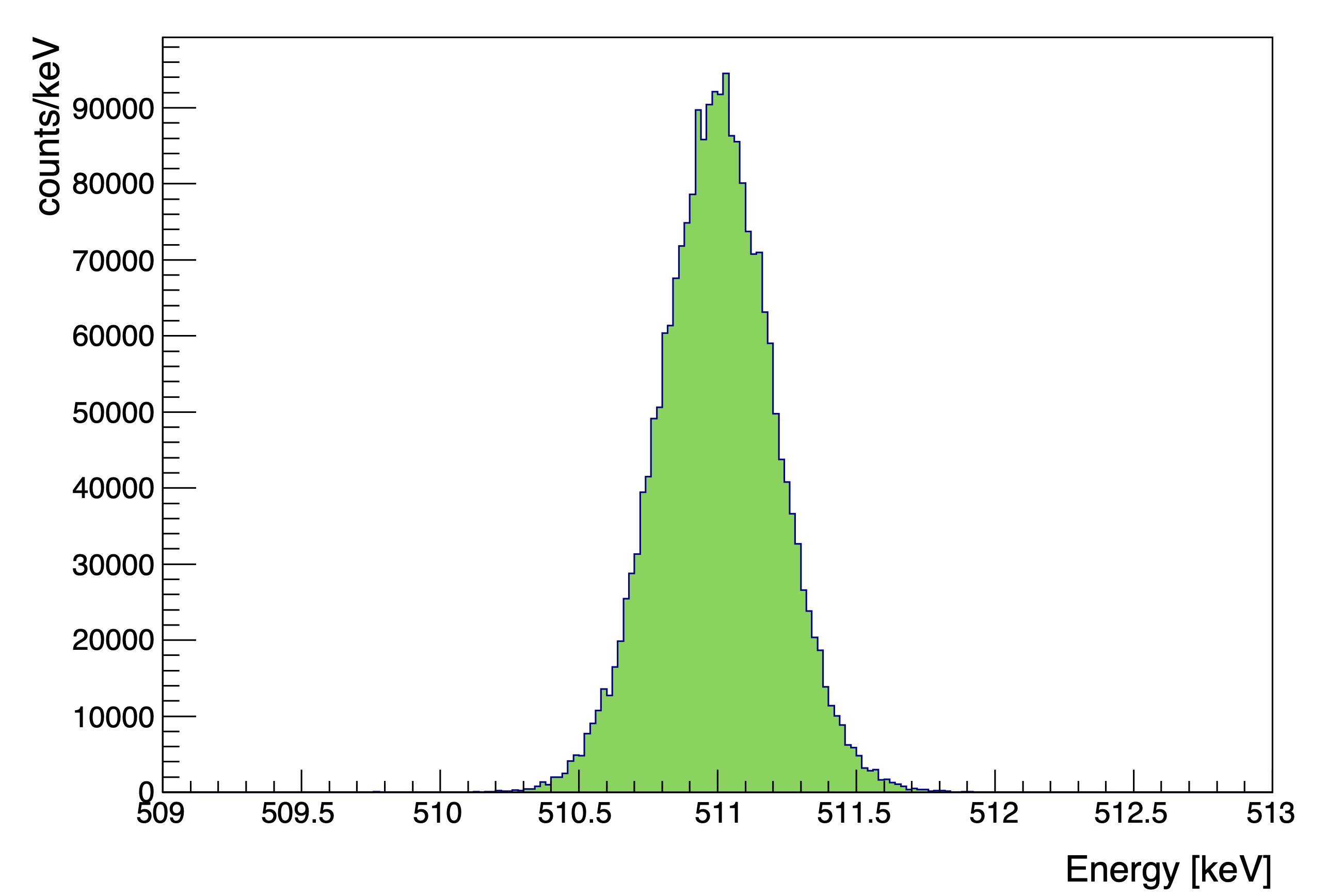}\quad%
  \caption{Energy spectrum of the simulated {\it 511-CAM}. The left panel shows the overall spectrum. The right panel focuses in on the 390\,eV FWHM 511\,keV peak. }
  \label{fig:spectru}
\end{figure}

These simulations allow us to determine the detector performance and optimize the instrument design with respect to the desired scientific objectives, including the effect of the 
gaps between absorbers.
The resulting 511\,keV energy spectrum is shown in Fig.\,\ref{fig:spectru}. The energy spectrum shows a pronounced peak centered on 511\,keV. The FWHM energy resolution is 390\,eV, and a total of 93\% of the detected 511\,keV events have reconstructed energies between 510.3\,keV and 511.8\,keV. The {\it 511-CAM} thus excels through its energy resolution: the {\it 511-CAM} energy resolution of 390\,eV is 11 times better than the 4.3\,keV resolution of COSI. Another strength of the {\it 511-CAM} is the 93\% efficiency 
of the detected events reconstructed within that energy resolution. 

We note that most photoelectric effect events deposit their energy in exactly one pixel. 
As noted above, these events have a particularly good energy resolution.
Even though one-pixel events are not suitable for a Compton imaging analysis
(which analyzes the energy deposited in two or more pixels, together with
the information about the location of these pixels), one-pixel events are 
expected to have a low background. 
For example, for the {\it X-Calibur} experiment, 
using only single-pixel events removes $>$99\% of the background. We can furthermore use self-shielding (i.e., discarding events triggering the edge pixels) to suppress one-pixel background events. For events with $\ge$\,2 triggered pixels, we can use Compton imaging to suppress events which cannot possibly come from the $\gamma$-ray optics \cite{2004SPIE.5540..144X,2020EPJWC.22506006D}. 

A possible design optimization is the use 
of ``staggered detector layers'' in which 
the layers are offset slightly to reduce the 
loss of photons that hit the gaps of the 
detector assembly. A full tradeoff study 
that accounts for the gain of detected photons 
but also evaluates the effect on the energy 
resolution is outside the scope of this paper.
\section{Predicted performance of a balloon-borne {\it 511-CAM} mission}
\label{s:performance}
\begin{figure}
  \centering
  \includegraphics[width=.48\textwidth]{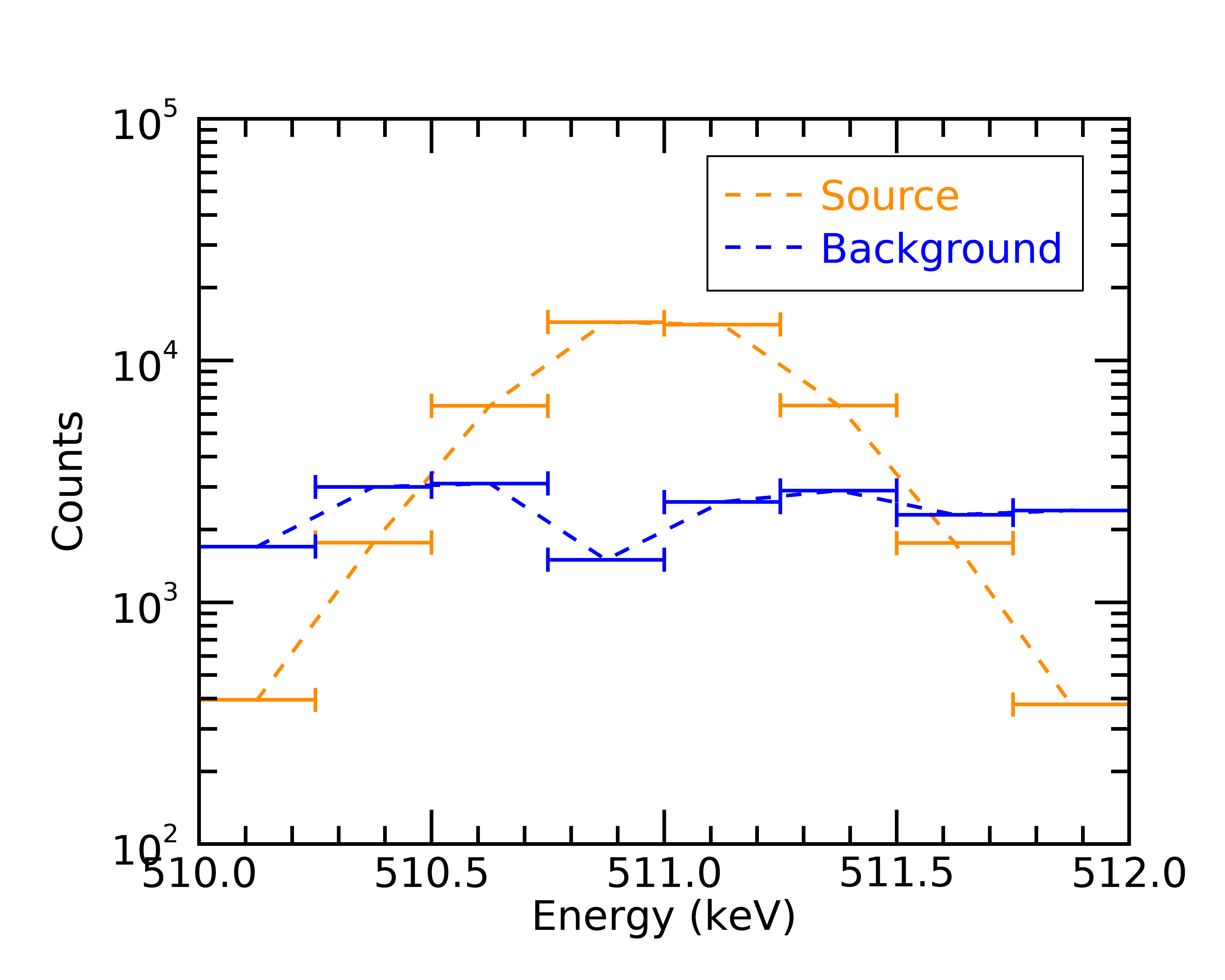}\quad%
  \includegraphics[width=.48\textwidth]{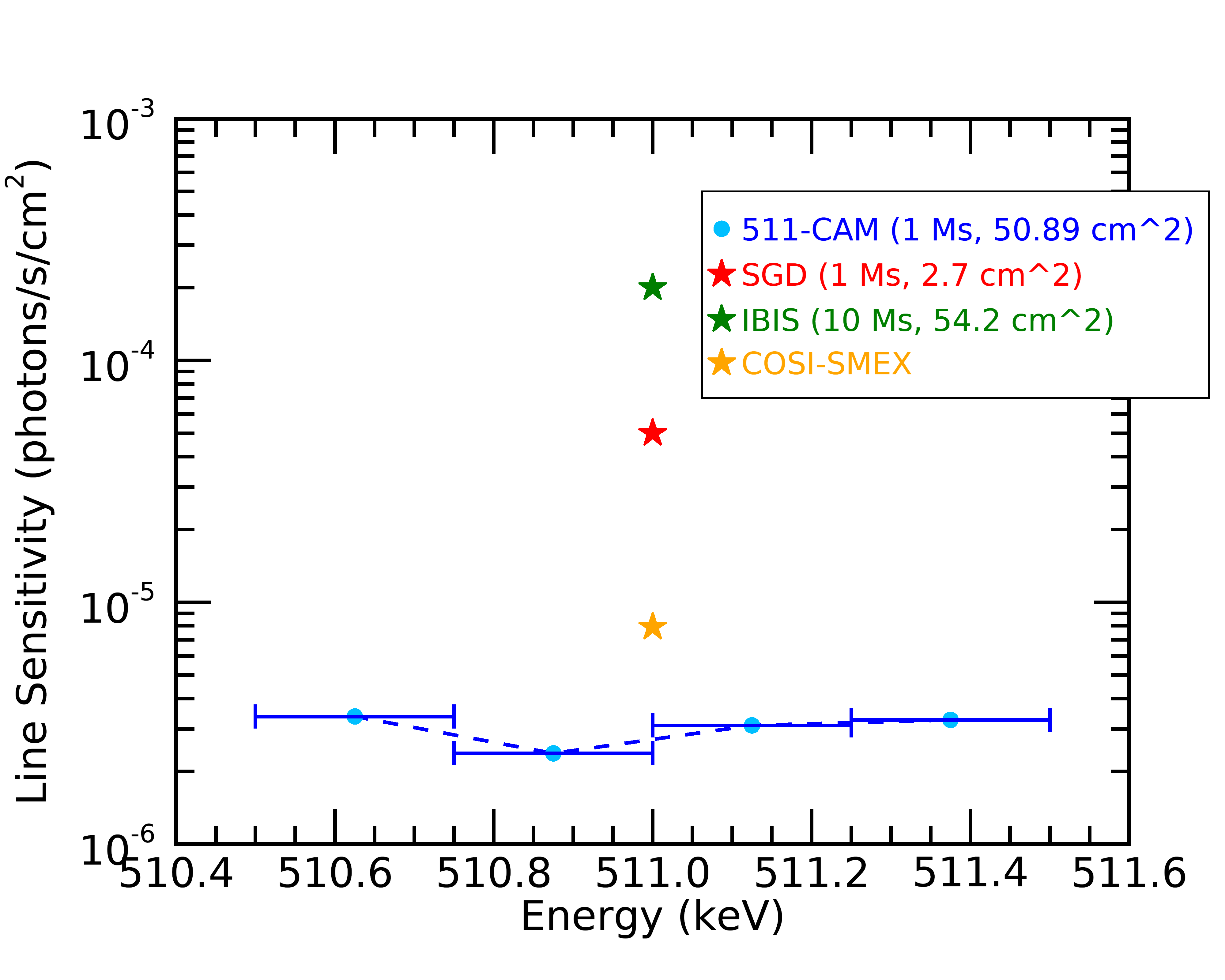}\quad%
  \caption{1\,Ms detected counts of simulated 511\,keV source and background of a balloon-borne mission (left panel). The 3$\sigma$ sensitivity comparison with line sensitivity of other instruments (right panel).}
  \label{fig:sensitivity}
\end{figure}

One of the key elements of the detector response is the instrument efficiency or effective area, which, for the {\it 511-CAM}, consists of both lens and detector efficiency. The simulation code calculates the lens channeling efficiency by using the relation
\begin{equation}
P(E) = {\rm Reflectance} \times {\rm Absorption} \times 
{\rm Open\,Fraction}.
\label{e:eff}
\end{equation}
The open fraction term is equivalent to the ratio of spacer layer thickness to the period of bi-layers. Our {\it 511-CAM} simulation shows that for the 9\,cm diameter lens, the resulting effective area of a balloon-borne mission at 511\,keV is {50.89\,cm$^2$}. To predict the sensitivity of the detector, we also modeled the dominant background components during a balloon flight and their effects on instrument performance. We followed the model developed by Gehrels \cite{GEHRELS1985324} for shield leakage of $\gamma$-ray background. The results are given for a column density of 3.5\,g/cm$^2$ of rest-atmosphere. Figure \ref{fig:sensitivity} shows the source and background counts (left panel) and the calculated sensitivity (right panel) at 3$\sigma$ confidence level. 
The estimated detector response under the concentrated point source and background indicates that {\it 511-CAM}, combined with the $\gamma$-ray concentrator, can achieve a sensitivity better than current instruments -- even for the most conservative estimate of the per-pixel energy resolution.
The development of SLEDGEHAMMER detectors with BiSn 
absorbers could give much better energy resolutions.
The background fluctuations scale with the square root of the
energy resolution. Thus, an energy resolution improved by
a factor of $\sim$\,4 would improve the sensitivity by a factor of 2. The key assumptions and results of these channeling optics/{\it 511-CAM} simulations are listed in Table \ref{tab:simulation}. 
\section{Discussion}
\label{s:discussion}
\begin{table}[t!] 
{\small
	\centering
	\begin{tabular}{ | p{7cm} | p{7cm} | }
		\hline
		\textbf{Parameter} & \textbf{Value}\\[5pt]
		\hline
		Number of channeling rings & $4$\\[5pt]
		\hline
	    Radius of rings & $2.25, 3, 3.75, 4.5$\,cm\\[5pt]
		\hline
		Multilayer bending angles & $0.11, 0.14, 0.18, 0.22^{\circ}$\\[5pt]
		\hline
	    Multilayer lengths & $4.6, 3.5, 2.7, 2.1$\,cm\\[5pt]
		\hline
	    Multilayer width & $\sim 1$\,cm\\[5pt]
		\hline
	    Total number of multilayers  & $\sim 83$\\[5pt]
		\hline
	    Focal length & {$12$\,m}\\[5pt]
		\hline
	    Focused beam diameter & $3.6$\,cm\\[5pt]
		\hline
	    Absorber material & $8\, \rm layers\ Bi$\\[5pt]
		\hline
	    Pixel volume & $\rm 1.45\times1.45\times2\,mm^3$\\[5pt]
		\hline
	    Pixel array & $\rm 20\ by\ 20$\\[5pt]
		\hline
		FWHM pixel resolution & $\rm 230\,eV\ @\ 511$\,keV\\[5pt]
		\hline
	    Trigger threshold & $1.2$\,keV\\[5pt]
		\hline
	    Observation duration & $1$\,Ms\\[5pt]
		\hline
		Flux & $0.1$\,ph/s\\[5pt]
		\hline
		FWHM energy resolution  & $\rm 390\,eV\ @\ 511$\,keV\\[5pt]
		\hline
        Channeling optics transmissivity & $80\%$\\[5pt]
		\hline
		 Effective area of optics & {$\rm 50.89\,cm^2$}\\[5pt]
		        \hline
		Detection efficiency of TES stack, incl.\ gaps & $65\%$\\[5pt]
		\hline
	\end{tabular}}
	\caption{Simulation parameters and simulation results for the combination of channeling optics and the {\it 511-CAM} focal plane detector. }
	\label{tab:simulation}
\end{table}
This paper discusses the new {\it 511-CAM} mission concept based on a stack of microcalorimetric 
detectors that combine $\sim$\,mm$^3$-sized absorbers with Transition Edge Sensors.  
The pointed mission would complement wide-field-of-view missions such as {\it COSI} 
\cite{2020ApJ...895...44K,2020ApJ...897...45S}, 
{\it COSI-X} \cite{2017ifs..confE.179Z}, or {\it AMEGO-X}
\cite{2022icrc.confE.649F}.
With the conservative assumption of using Bi absorbers, the {\it 511-CAM} mission would achieve a per-pixel energy deposit resolution of 230\,eV\,FWHM and a 511\,keV gamma-ray energy resolution of 390\,eV\,FWHM.
We argue that high-Z absorbers with lower heat capacities, such as the Bi$_{0.57}$-Sn$_{0.43}$ alloy, 
could lead to further marked improvements in energy resolution.
The energy resolution of a {\it 511-CAM}-type mission directly impacts the sensitivity of a 
background-limited 511\,keV gamma-ray mission.

The proposed balloon-borne mission would be ideally suited to detect point sources of 511 keV emission, offering improved localization of these sources and precise line-of-sight velocity measurements. If the 511 keV emission from the galactic center region is truly diffuse, the mission could still detect it, 
but the estimated detection rate would be low, i.e., 0.17 photons per day over a background of 0.03 photons per day.

A mission with a broad energy bandpass might be more attractive than a mission focusing exclusively on the detection of the 
511 keV line. The channeling optics discussed above 
inherently come with a broad bandpass, thus enabling the search
for gamma-ray lines from various astrophysical sources. 
Alternatively, a broadband mission could be built by combining 
a re-configurable Laue lens \cite{2021ExA....51..153L} 
with an extendable optical bench. One could envision 
multiple flights of such a mission, each one using
a lens and focal length configuration optimized for a certain
science goal.
\section*{Acknowledgements}
HK thanks N. Barrière for 
the permission to use Fig.\ 1, 
and NASA for the support through the grants NNX16AC42G and 80NSSC18K0264.
The authors acknowledge the helpful comments of two anonymous referees.
\bibliography{paper511}
\bibliographystyle{spiejour} 
\end{document}